\documentclass[fleqn,usenatbib]{mnras}

\usepackage{newtxtext,newtxmath}
\usepackage[T1]{fontenc}

\DeclareRobustCommand{\VAN}[3]{#2}
\let\VANthebibliography\thebibliography
\def\thebibliography{\DeclareRobustCommand{\VAN}[3]{##3}\VANthebibliography}

\usepackage{graphicx}	% Including figure files
\usepackage{amsmath}	% Advanced maths commands
\usepackage{gensymb}
\usepackage{orcidlink}
\usepackage{color} %Just for markup while writing
\usepackage{hyperref}
\usepackage{cleveref}
\usepackage{verbatim}

\newcommand{\kms}{\mathrm{km\,s^{-1}}}
\newcommand{\Msun}{\mathrm{M_\odot}}
\newcommand{\lowspec}{\textit{LowSpec}}
\newcommand{\highspec}{\textit{HighSpec}}
\newcommand{\highark}{\textit{HighSpecArk}}
\newcommand{\medark}{\textit{MedSpecArk}}
\newcommand{\lhighspec}{\textit{LHighSpec}}
\newcommand{\lhighark}{\textit{LHighSpecArk}}
\newcommand{\lhigharkref}{\textit{LHighSpecArkRef}}
\newcommand{\subfive}{fiducial}

\defcitealias{Chevalier1985}{CC85}
\defcitealias{Springel2003}{SH03}

\crefname{section}{Section}{Sections}
\Crefname{section}{Section}{Sections}
\crefname{subsection}{Section}{Sections}
\Crefname{subsection}{Section}{Sections}
\crefname{subsubsection}{Section}{Sections}
\Crefname{subsubsection}{Section}{Sections}
\crefname{figure}{Fig.}{Fig.}
\Crefname{figure}{Fig.}{Fig.}
\crefname{equation}{equation}{equations}
\Crefname{equation}{Equation}{Equations}
\creflabelformat{equation}{#2#1#3}
\crefname{table}{Table}{Tables}
\Crefname{table}{Table}{Tables}
\crefname{appendix}{Appendix}{Appendices}
\Crefname{appendix}{Appendix}{Appendices}
\title[Arkenstone I]{Arkenstone I: A Novel Method for Robustly Capturing High Specific Energy Outflows In Cosmological Simulations}

\author[M. C. Smith et al.]{
Matthew C. Smith\orcidlink{0000-0002-9849-877X},$^{1,2,3}$\thanks{E-mail: msmith@mpa-garching.mpg.de}
Drummond B. Fielding\orcidlink{0000-0003-3806-8548},$^{4}$
Greg L. Bryan\orcidlink{0000-0003-2630-9228},$^{5,4}$
Chang-Goo Kim\orcidlink{0000-0003-2896-3725},$^{6}$
\newauthor
\ Eve C. Ostriker\orcidlink{0000-0002-0509-9113},$^{6}$
Rachel S. Somerville\orcidlink{0000-0003-2835-8533},$^{4}$
Jonathan Stern\orcidlink{0000-0002-7541-9565},$^{7}$
Kung-Yi Su\orcidlink{0000-0003-1598-0083},$^{8}$
Rainer Weinberger\orcidlink{0000-0001-6260-9709},$^{9}$
\newauthor
\ Chia-Yu Hu\orcidlink{0000-0002-9235-3529},$^{10,11}$
John C. Forbes\orcidlink{0000-0002-1975-4449},$^{4,12}$
Lars Hernquist\orcidlink{0000-0001-6950-1629},$^{13}$
Blakesley Burkhart\orcidlink{0000-0001-5817-5944}$^{14,4}$
and Yuan Li\orcidlink{0000-0001-5262-6150}$^{15}$
\\
$^{1}$Max-Planck-Institut f{\"u}r Astrophysik, Karl-Schwarzschild-Str. 1, D-85748, Garching, Germany\\
$^{2}$Universit{\"a}t Heidelberg, Zentrum f{\"u}r Astronomie, Institut f{\"u}r Theoretische Astrophysik, Albert-Ueberle-Str. 2, D-69120 Heidelberg, Germany\\
$^{3}$Max-Planck-Institut f{\"u}r Astronomie, K{\"o}nigstuhl 17, D-69117 Heidelberg, Germany\\
$^{4}$Center for Computational Astrophysics, Flatiron Institute, 162 5\textsuperscript{th} Avenue, New York, NY 10010, USA\\
$^{5}$Department of Astronomy, Columbia University, 550 West 120\textsuperscript{th} Street, New York, NY 10027, USA\\
$^{6}$Department of Astrophysical Sciences, Princeton University, 4 Ivy Lane, Princeton, NJ 08544, USA\\
$^{7}$School of Physics {\&} Astronomy, Tel Aviv University, Tel Aviv 69978, Israel\\
$^{8}$Black Hole Initiative, Harvard University, 20 Garden St., Cambridge, MA 02138, USA\\
$^{9}$Canadian Institute for Theoretical Astrophysics, 60 St. George Street, Toronto, ON M5S 3H8, Canada\\
$^{10}$Max-Planck-Institut f\"{u}r Extraterrestrische Physik, Giessenbachstrasse 1, D-85748 Garching, Germany\\
$^{11}$Department of Astronomy, University of Florida, 211 Bryant Space Science Center, Gainesville, FL 32611, USA\\
$^{12}$School of Physical and Chemical Sciences|Te Kura Mat\={u}, University of Canterbury, Private Bag 4800, Christchurch 8140, New Zealand\\
$^{13}$Harvard-Smithsonian Center for Astrophysics, 60 Garden Street, Cambridge, MA 02138, USA\\
$^{14}$Department of Physics and Astronomy, Rutgers University, 136 Frelinghuysen Rd, Piscataway, NJ 08854, USA\\
$^{15}$Department of Physics, University of North Texas, Denton, TX 76203, USA
}

\date{Accepted XXX. Received YYY; in original form ZZZ}

\pubyear{2023}

\begin{document}
\label{firstpage}
\pagerange{\pageref{firstpage}--\pageref{lastpage}}
\maketitle

\begin{abstract}
\textsc{Arkenstone} is a new model for multiphase, stellar feedback-driven galactic winds designed for inclusion in coarse resolution cosmological simulations. In this first paper of a series, we describe the features that allow \textsc{Arkenstone} to properly treat high specific energy wind components and demonstrate them using idealised non-cosmological simulations of a galaxy with a realistic circumgalactic medium (CGM), using the \textsc{Arepo} code. Hot, fast gas phases with low mass loadings are predicted to dominate the energy content of multiphase outflows. In order to treat the huge dynamic range of spatial scales involved in cosmological galaxy formation at feasible computational expense,
cosmological volume simulations typically employ a Lagrangian code or else use adaptive mesh refinement with a quasi-Lagrangian refinement strategy.
However, it is difficult to inject a high specific energy wind in a Lagrangian scheme without incurring artificial burstiness. Additionally, the low densities inherent to this type of flow result in poor spatial resolution. \textsc{Arkenstone} addresses these issues with a novel scheme for coupling energy into the transition region between
the interstellar medium (ISM) and the CGM, while also providing refinement at the base of the wind. Without our improvements, we show that poor spatial resolution near the sonic point of a hot, fast outflow leads to an underestimation of gas acceleration as the wind propagates. We explore the different mechanisms by which low and high specific energy winds can regulate the star formation rate of galaxies. In future work, we will demonstrate other aspects of the \textsc{Arkenstone} model.
\end{abstract}


%%%%%%%%%%%%%%%%%%%%%%%%%%%%%%%%%%%%%%%%%%%%%%%%%%

%%%%%%%%%%%%%%%%% BODY OF PAPER %%%%%%%%%%%%%%%%%%

\section{Introduction} \label{sec:intro}
A fundamental component of galaxy formation, in the $\Lambda$CDM
cosmological framework, is the radiative cooling and subsequent
accretion of gas into the centre of the gravitational potential
of dark matter haloes where it can form stars \citep{White1978,White1991}.
However, simulations that only include the physics of gravity, hydrodynamics and radiative cooling
fail to reproduce realistic galaxies, with an overproduction of stars in highly compact,
massive discs \citep{Katz1991,Navarro1991,Katz1992,Navarro1993,
Navarro1994,Navarro1995}. 
The simple hypothesis that stars should form in the interstellar medium (ISM) of galaxies
on the order of the dynamical time underestimates the true timescale for star formation
by a factor of 20 - 100 \citep{Zuckerman1974,Williams1997,Kennicutt1998,
Evans1999,Krumholz2007,Evans2009, Utomo2018}. In addition to an overly efficient conversion of
gas to stars, neglecting some additional form of energy injection
leads to predicted overall galaxy baryon fractions that are far in excess of observations
\citep[e.g.][]{White1991,Keres2009}. Furthermore, observations of the circumgalactic
medium of galaxies (CGM) show metal enrichment, requiring that mass processed through
stars must be thrown back out into the halo \citep[e.g.][]{Aguirre2001,Pettini2003,
Songaila2005,Songaila2006,Martin2010}. 
Complementary to the above galactic-scale constraints, it has long been understood 
that inputs of energy are needed to maintain the observed thermal and 
turbulent properties of the ISM of the Milky Way and other spiral galaxies 
\citep[e.g.][]{Wolfire1995,Wolfire2003,Ostriker2022}

Significant progress has been made towards resolving these discrepancies by invoking
feedback processes from massive stars and active galactic nuclei (AGN)
\citep[see e.g. reviews by][]{Somerville2015,Naab2017}. 
Canonically, stellar feedback is thought to be the dominant mechanism in lower mass
haloes while AGN feedback operates in more massive systems.
With reference
to the problems described above, feedback has three main impacts:
\begin{enumerate}
\item A local action, driving turbulence, heating the ISM and reducing the efficiency per dynamical time with which ISM material is converted
to stars.
\item The ejection of gas from the ISM, reducing the available reservoir of star
forming material and enriching the CGM.
\item The prevention of fresh inflows of gas from reaching the galaxy from the intergalactic medium (IGM) via the CGM.
\end{enumerate}

The latter two items manifest in the form of galactic winds. Galactic winds are
observed across the full span of cosmic time in a broad range of star forming galaxies \citep[e.g.,][]{Heckman1990, Martin1999, Shapley2003, Weiner2009, Rubin2014}. A fundamental aspect of observed galactic winds is their multiphase nature. They are commonly observed in both emission and absorption with tracers that probe gas at $\lesssim 100$K \citep[e.g.,][]{Rupke2005,Bolatto2013,Martini2018}, $\sim 10^4$K \citep[e.g.,][]{Martin2009,Westmoquette2009,Nielsen2015}, $\sim 10^{5.5}$K \citep[e.g.,][]{Steidel2010,Kacprzak2015,Chisholm2018}, and $\gtrsim 10^7$K \citep[e.g.,][]{Strickland2009,Lopez2020,Hodges-Kluck2020}.

Stellar feedback driven winds are primarily powered by energy injection from core-collapse
supernovae (SNe).
Early analytic models of galactic winds described high specific energy wind components
i.e. hot, fast, low density flows. \cite{Chevalier1985} (hereafter \citetalias{Chevalier1985})
models a wind powered by the injection of mass and energy into a spherically symmetric,
finite region. Extensions to this basic model have added the effect of radiative cooling and gravity \citep[e.g.,][]{Wang1995, Thompson2016}, more realistic injection \citep{Bustard2016, Nguyen2023}, and spatially extended mass-loading and non-spherical expansion \citep{Nguyen2021}. Recently, it has been shown that including the two-way interaction between the hot, volume-filling phase and the cold, clumpy phase is essential for explaining key characteristics of galactic winds \citep{Fielding2022}.

The generation of winds by stellar feedback within the ISM is 
sensitive to a variety of small-scale effects. These have been studied with the aid of 
high resolution simulations of $\sim$kpc patches of the ISM. These have revealed that the
efficient coupling of SN energy into galactic winds is dependent on the placement
of SNe \citep[e.g.][]{Creasey2013,Martizzi2016,Girichidis2016a,Li2017} and the
degree of SN clustering in space and time \citep[e.g.][]{Kim2017b,Fielding2018}.
Recent ISM patch simulations have included self-consistent formation of stars
and a range of stellar feedback channels \citep[e.g.][]{Kim2017a,Gatto2017,Kim2018,
Rathjen2021}. An important feature of these simulations is the production of outflows
that are multiphase as they leave the ISM. \cite{Kim2020b,Kim2020a} show that
energy and mass are not partitioned evenly between the phases; a hot ($\gtrsim 10^6$K), fast component 
dominates the energy loading while a cool ($\lesssim 10^4$K), slow component dominates the mass loading.
The collection of the majority of available wind power into a high specific energy
wind component appears to be a ubiquitous phenomenon \citep[see e.g. the
compilation of][]{Li2020a}.

Despite their utility, ISM patch simulations are limited by their idealised nature.
They lack both the spatial extent and the correct geometry to follow the subsequent
evolution of the wind as it travels out into the CGM \citep[][]{Martizzi2016}.
Global simulations of individual galaxies alleviate these issues, but the resolution requirements
typically limit studies to experiments with
highly idealized implementations of star formation and stellar feedback
\citep[e.g.][]{Tanner2016,Fielding2017,Schneider2018,Schneider2020}
or to extremely low mass galaxies \citep[e.g.][]{Hu2017,Smith2018,Emerick2018,Hu2019,Gutcke2021,Smith2021a,Smith2021b,Hislop2022,Andersson2023,Steinwandel2023}.
To gain the
greatest insight into the complexities of galaxy formation, we must simulate systems
of a wide range of masses, merger histories and environments in a cosmological context. 
Even in a narrow halo mass range, CGMs drawn from the same cosmological simulation can
display a significant diversity of properties \citep[e.g.][]{Ramesh2023}.

However, simulations of cosmological
volumes lack the necessary resolution to treat key physical processes in wind driving
from first principles. 
Highly abstracted subgrid prescriptions must be adopted that 
attempt to compensate for the lack of a properly resolved ISM.
Some cosmological volume simulations use models that inject SNe energy and/or momentum directly into the 
ISM \citep[e.g.][]{Dubois2014,Schaye2015,McCarthy2017,Feldmann2023}.
Typically, lack of mass resolution
imposes a minimum mass into which the energy can be injected, resulting 
in a low temperature increase and short cooling time if the energy from individual
SN events is separately dumped into the gas.
With a short cooling time, energy cannot build up over multiple events to drive an efficient outflow.
At low mass resolution, it is impossible to create a high specific
energy, low mass loaded component such as that seen in resolved simulations. The
required temperature increase can be achieved via numerical sleight of hand,
discretising available feedback energy into rare, high magnitude injections
of energy \citep[e.g. the stochastic heating scheme of][]{DallaVecchia2012}.
In principle, direct injection of energy into the ISM has the advantage that it can promote
some of the local and
long-range impacts of feedback simultaneously. However,
the lack of resolution and the forced clustering of energy means that the feedback-ISM
interaction can only be qualitatively correct, at best. The driving of small-scale
turbulence by feedback cannot be properly captured, neither the generation of
a multiphase and porous ISM structure \citep[see e.g.][]{Kim2017a}.
While a multiphase outflow
may emerge, its properties are unlikely to comport with those seen in simulations
that properly resolve the generation of the wind within the ISM
\citep[e.g.][show that $\lesssim100\,\mathrm{M_\odot}$ resolution is required to achieve convergent wind properties]{Smith2018,Hu2019}.
Therefore, schemes that inject feedback directly into the ISM at coarse resolution, whether they involve directly tunable parameters or not, must
be viewed as effective models. 

Alternatively, some cosmological volume simulations avoid treating the feedback-ISM interaction 
explicitly,
instead adopting non-local source terms to seed the wind in the ISM/CGM transition region
\citep[e.g.][]{Vogelsberger2014,Dave2016,Pillepich2018,Henden2018,Dave2019,Pakmor2023}.
These source terms are typically mediated with the use of hydro-decoupled wind particles
\citep{Springel2003}.
These are spawned from star forming gas and travel through the ISM without participating
in hydrodynamic interactions until they have recoupled their mass, energy, momentum and
metals into the edge of the CGM. This recoupling usually involves a simple dumping
of wind particle conserved quantities into the ambient medium \citep[an exception being
the scheme of][which assumes all wind particles are clouds and shreds them
gradually into the ambient medium]{Huang2020}.
The missing impact of feedback on the small-scale ISM is then included via some other
subgrid model, typically through the use of an effective equation of state
\citep[e.g.][]{Springel2003}.
Such schemes clearly utilise numerical slight of hand to drive outflows
and regulate the ISM. However, it should be pointed out that schemes that do not hydro-decouple
outflows when they lack the resolution to properly resolve their passage out
of the ISM are at least as unphysical.
The use of a non-local source term approach grants finer 
control over the injection of the wind, removing the reliance on producing emergent properties of 
the outflow through an under-resolved ISM.

Conventionally, wind particle schemes launch one wind component,
with a single velocity, temperature and mass loading \citep[an exception being][]{Dave2016}
that is dependent on some large scale galaxy or halo property (e.g. stellar mass,
dark matter velocity dispersion etc.).
Thus, unlike the case in resolved simulations, their outflow is typically single phase at launch
and the outflow properties do not reflect local galactic conditions.
While the vast majority of schemes do not inject multiphase winds,
the highest resolution cosmological simulations do form a multiphase structure as they 
propagate out into the
CGM \citep[e.g.][]{Nelson2019,Mitchell2020,Pandya2021}. However, even these lack the
resolution to reliably resolve the interaction between the cooler wind phases
and the hot ambient flow. Specifically, to even roughly capture the mass balance between the hottest and coldest phases requires resolving the largest eddies in the cloud mixing layers.
This corresponds to resolving the clouds by at least 16 cells in diameter (e.g., resolving a $T=10^4\,$K, $M_{\rm cloud} = 10^4\,\mathrm{M_\odot}$ cloud in a wind with $P=10^3 k_B {\rm K \, cm}^{-3}$ requires $\Delta x \lesssim$ 10 pc, or a mass resolution of $m_{\rm cell} \lesssim 1 \, M_\odot$; e.g., \citealt{Gronke2020,Gronke2022}). Accurately capturing the morphology or detailed phase structure demands even more stringent resolution requirements \citep{Abruzzo2022b}.
Another important issue is that the mass and energy loading factors (or equivalents)
that are used in most models (either directly as input parameters or 
as a consequence of model design) are typically far higher
than those measured in smaller scale resolved simulations \citep[see e.g.][]{Li2020a}.
This implies that either other important physical mechanisms are missing
or that some feature of the design of subgrid wind schemes
force these unphysically high loadings.
Finally, we note that while the varied galaxy formation models
deployed in contemporary simulations
are capable of creating galaxies with realistic stellar components,
they can do so in very different ways, resulting in disparate halo gas properties \citep{Davies2020,Kelly2022,Ayromlou2023}.

The SMAUG (Simulating Multiscale Astrophysics to Understand Galaxies)
project\footnote{\url{https://www.simonsfoundation.org/flatiron/center-for-computational-astrophysics/galaxy-formation/smaug}} 
was motivated by the need to create large volume cosmological simulations with realistic and predictive treatments of galaxy formation processes. For the reasons just discussed, all current cosmological simulations of galaxy formation are forced to adopt subgrid recipes to treat key physical processes, including star formation, stellar feedback, and black hole seeding, growth, and feedback, among others. A common practice is to tune the parameters that govern these phenomenological subgrid recipes such that a set of chosen observations is reproduced. This approach clearly has a number of drawbacks, including a reduction in predictive power. Additionally, numerical simulations are in general too expensive to thoroughly explore parameter degeneracies. And as alluded to above, different subgrid implementations currently make very different predictions for quantities that have not been explicitly calibrated. The goal of SMAUG is to use multi-scale simulations that probe the relevant physics on the scales where it operates, augmented by analytic models, to develop physically grounded subgrid recipes that no longer require phenomenological tuning. For example, \citet{Kim2020b,Kim2020a} measured the emergent properties of multiphase stellar driven winds in resolved ISM simulations and parameterized them based on local galaxy properties, with the explicit goal of providing input to larger scale cosmological simulations. 

\textsc{Arkenstone}, a core component of the SMAUG project, is a new subgrid model for multiphase winds. 
\textsc{Arkenstone} is designed for simulations
where the ISM cannot be properly resolved and is intended for eventual
use in large volume cosmological simulations. 
In order to have a fine
control over the injection of the wind, in the face of a wide gamut of
important yet unresolvable physics, \textsc{Arkenstone} makes use of a
scheme that employs wind particle propagation and recoupling at a location different from the launch point.
The full \textsc{Arkenstone} model has three novel features: 
\begin{enumerate}
\item Winds are
    launched with hot and cool components with separate mass and energy loadings.
    Within these components, wind temperature and velocity are drawn from
    distributions with parameters calibrated in resolved simulations \citep[e.g.][]{Kim2020a}.
\item Because this splitting of the wind into two phases results
    in a high specific energy component, special attention has to be paid
    to the way in which these hot, fast, low density flows are resolved.
\item We utilise a new ``cloud particle'' scheme to treat cold
    clouds embedded in the hot flow. These exchange mass, energy, momentum
    and metals bidirectionally with the hot flow.
    Clouds lose mass as they are shredded by the interaction with the ambient
    medium, but can also gain mass as hot wind material cools onto them,
    providing a significant source of acceleration in some cases
    \citep[see e.g.][]{Gronke2018,Gronke2020,Schneider2020,Gronke2022,Fielding2022}.
    This contrasts with the
    scheme of \cite{Huang2020}, in which clouds can only lose mass and momentum and
    which does not include a separately injected hot wind component.
\end{enumerate}
In this first presentation of our work, we describe and demonstrate the second
of the feature sets described above. 
In \cref{sec:challenge}, we elucidate the challenges inherent
to simulating high specific energy galactic winds with coarse resolution
and pseudo-Lagrangian schemes.
In \cref{sec:methods}, we describe the relevant numerical details of
the \textsc{Arkenstone} model.
In \cref{sec:results}, we
use non-cosmological idealised 
simulations of
galaxies with a realistic CGM, to show how
\textsc{Arkenstone} is able to solve the problems outlined
in \cref{sec:challenge}. Our fiducial simulations
are presented in \cref{sec:fid_sims}.
We also explore the
different ways in which low and high specific energy winds can
regulate galaxy properties in this section. 
Following on from our fiducial demonstration, 
we examine the sensitivity of results to assumptions about wind launch
direction (\cref{sec:direction}),
show the model's resolution
dependence (\cref{sec:resolution}),
study how the model
behaves with a different CGM configuration (\cref{sec:supersonicICs})
and explore a less energy loaded wind (\cref{sec:twindfix}).
In \cref{sec:discussion} we discuss our findings.
In future work,
we will describe the remaining aspects of the \textsc{Arkenstone}
model. 

\section{The challenges of simulating high specific energy winds} \label{sec:challenge}
The specific energy of a wind is the energy per unit mass carried by the wind. It is
therefore the ratio of the energy flow rate, $\dot{E}$, to the mass flow rate,
$\dot{M}$. Increasing the specific energy of the wind necessarily means increasing
the temperature and/or velocity of the wind. It is frequently convenient to
characterise winds in terms of their specific energy content rather than their
temperature or velocity because the fraction of the total energy carried in
thermal or kinetic components may change as the flow progresses.
Lagrangian hydrodynamic schemes maintain constant mass per resolution element.
Additionally, Eulerian schemes employing adaptive mesh refinement (AMR)
may also use refinement criteria in cosmological simulations
which function in a similar manner.
When simulating high specific energy winds, this can pose a challenge to
temporal and spatial resolution, as we now describe.

\subsection{Temporal resolution} \label{sec:temporal_res}
As described in \cref{sec:intro}, schemes for injecting feedback energy in
cosmological simulations often do so in discrete events. Often, the scheme
mandates a particular specific energy, $e_\mathrm{inj}$, associated with an injection
event. This may be the velocity and thermal energy of a wind particle
\cite[e.g.][]{Springel2003} or the minimum temperature that a portion of the ISM
must be raised to \cite[e.g.][]{DallaVecchia2012}. The net energy injected in an
event is $E_\mathrm{inj}=e_\mathrm{inj}m_\mathrm{inj}$, where $m_\mathrm{inj}$ is
a characteristic mass associated with the injection. $m_\mathrm{inj}$ may correspond
to the mass of a wind particle or the mass of a certain number of ISM gas resolution
elements. In either case, it is typically a numerical parameter rather than
corresponding to anything physical. Therefore, the energy resolution of the wind injection,
$E_\mathrm{inj}$, is tied to the mass resolution. This means that as the specific energy
of the injected wind is increased, the injection becomes increasingly noisy in time
at fixed mass resolution. Likewise, at fixed specific energy the energy injection
is divided into rarer, more energetic events as the mass resolution is coarsened.
The issue of poor temporal resolution is exacerbated if the injected wind temperature
and velocity are not single valued, but are instead drawn from a distribution
\citep[e.g. the \textsc{Twind} scheme of][implemented in \textsc{Arkenstone}
but not used in this paper]{Kim2020a}.
In this case, the mass resolution must be even higher
in order to properly sample the distribution.

\subsection{Spatial resolution} \label{sec:spatial_res}
\begin{figure} 
\centering
\includegraphics{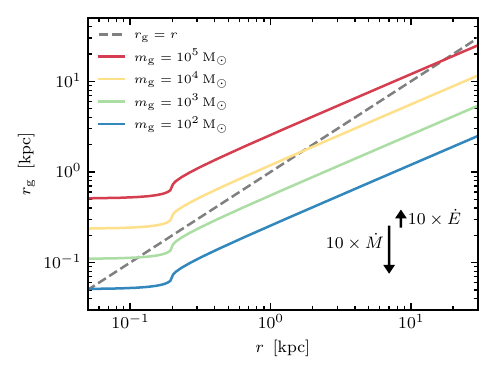}
\caption{The spatial resolution as a function of distance, approximating mass resolution elements
as spheres with radius $r_\mathrm{g}$, for a \citetalias{Chevalier1985} wind solution with
$r_\mathrm{inj}=200\,\mathrm{pc}$,
$\dot{M}=0.1\,\mathrm{M_\odot}\,\mathrm{yr^{-1}}$ and $\dot{E}=10^{42}\,\mathrm{erg}\,\mathrm{s^{-1}}$.
Different coloured lines correspond to different mass resolutions. The dashed line indicates
$r_\mathrm{g}=r$. Solutions that fall above this line indicate a completely spatially
unresolved wind. The arrows
indicate how the normalisation of the solution for $r \gg r_\mathrm{inj}$ would change if $\dot{M}$ or $\dot{E}$ were increased by a
factor of 10 (there is no dependency on $r_\mathrm{inj}$).
This demonstrates that high specific energy winds require fine mass resolution, particularly at the base of the wind.\vspace{-4ex}}
\label{fig:cc85_resolution} 
\end{figure}

In Lagrangian (or quasi-Lagrangian) numerical methods,
spatial resolution coarsens as density decreases. This can make it difficult to
properly resolve the evolution of a high specific energy wind. To illustrate this
point in concrete terms, we will examine the analytic wind solution of \citetalias{Chevalier1985}.
This steady state and spherically symmetric solution applies to an adiabatic wind generated by the constant 
and spatially uniform injection of mass and energy into a sphere of radius $r_\mathrm{inj}$,
ignoring gravity. The energy is injected into the sphere as a purely thermal component (i.e.
no momentum is injected) but the wind is accelerated out of the injection region, becoming
supersonic at $r_\mathrm{inj}$.

We can now examine how well a Lagrangian code would
spatially resolve such an outflow. We can approximate the gas resolution element as a sphere,
meaning that its radius varies with density as $r_\mathrm{g}=\left(3m_\mathrm{g}/4\pi\rho\right)^{1/3}$,
where its mass is $m_\mathrm{g}$. The asymptotic limit, far from $r_\mathrm{inj}$, of the density profile in the \citetalias{Chevalier1985} solution is
% \begin{align}
% \rho(r) &= 9.362\times10^{-29} \, {\rm g \, cm}^{-3}\nonumber \\
%         & \qquad \times \left( \frac{\dot{M}}{0.1 M_\odot \, {\rm yr}^{-1}}\right)^{\frac{3}{2}} \left( \frac{\dot{E}}{10^{42} \, {\rm erg \, s}^{-1}}\right)^{-\frac{1}{2}} \left( \frac{r}{1 \, {\rm kpc}} \right)^{-2}.
% \end{align}
\begin{align}
\rho(r) &= 9.362\times10^{-29} \, {\rm g \, cm}^{-3} \dot{M}_{0.1}^{\frac{3}{2}} \dot{E}_{42}^{-\frac{1}{2}} \, r_{\rm kpc}^{-2},
\end{align}
where $\dot{M}_{0.1}~=~\dot{M} / {0.1 M_\odot \, {\rm yr}^{-1}}$, $\dot{E}_{42}~=~\dot{E}/{10^{42} \, {\rm erg \, s}^{-1}}$, and $r_{\rm kpc}~=~r / {\rm kpc}$.
We can then obtain
an expression for $r_\mathrm{g}$ at a given distance:
% \begin{equation}
% \frac{r_\mathrm{g}}{r} \simeq 1.62 m_\mathrm{g}^\frac{1}{3} \dot{M}^{-\frac{1}{2}} \dot{E}^{\frac{1}{6}} r^{-\frac{1}{3}}. \label{eq:cc85_resolution}
% \end{equation}
\begin{equation}
\frac{r_\mathrm{g}}{r} \simeq 1.20 \, m_\mathrm{g,4}^\frac{1}{3} \dot{M}_{0.1}^{-\frac{1}{2}} \dot{E}_{42}^{\frac{1}{6}} r_{\rm kpc}^{-\frac{1}{3}}, \label{eq:cc85_resolution}
\end{equation}
where $m_\mathrm{g,4} = m_\mathrm{g} / 10^4 \mathrm{M_\odot}$.
It can therefore be seen, as expected, that increasing the mass per resolution element or increasing the specific
energy of the wind coarsens the spatial resolution. A similar derivation can be made to obtain
the spatial resolution at $r_\mathrm{inj}$ (which is also the sonic radius); this yields
the same form as \cref{eq:cc85_resolution} but with a prefactor 0.79 times smaller.

In \cref{fig:cc85_resolution}, we plot $r_\mathrm{g}$ as a function of $r$ for the full
\citetalias{Chevalier1985} solution for a wind with $r_\mathrm{inj}=200\,\mathrm{pc}$,
$\dot{M}=0.1\,\mathrm{M_\odot}\,\mathrm{yr^{-1}}$ and $\dot{E}=10^{42}\,\mathrm{erg}\,\mathrm{s^{-1}}$,
one of the set of parameters they suggest for the galaxy M82 (the weakest wind parameters
they examine). Different coloured lines indicate different mass resolutions. 
We indicate with
arrows how the normalisation of the solution for $r \gg r_\mathrm{inj}$ would change if $\dot{M}$ or $\dot{E}$ were increased by a
factor of 10. Note that for $r \gg r_\mathrm{inj}$ there is no dependence on the value of $r_\mathrm{inj}$.
We also plot the line $r_\mathrm{g}=r$ for reference. If a solution for a given $m_\mathrm{g}$ lies above
the $r_\mathrm{g}=r$ line, then the wind is completely spatially unresolved. In other words, the
galactocentric radius is resolved by fewer than one resolution element. It is clear that
at $10^{5}\,\Msun$ resolution (already towards the higher resolutions employed in contemporary
cosmological volume simulations), a \citetalias{Chevalier1985} wind with these parameters
would be significantly unresolved. A factor of 100 better mass resolution is required to
begin to resolve the evolution of the wind as it flows outwards from the injection region.
As is apparent from \cref{eq:cc85_resolution}, more mass loaded, lower specific energy winds
are easier to spatially resolve than less mass loaded, higher specific energy winds.

The simplifications made in the setup of the \citetalias{Chevalier1985} solution mean that it
not entirely applicable to the evolution of galactic winds in a more realistic galactic
context. \Cref{eq:cc85_resolution} essentially represents the worst case scenario,
since a wind propagating outwards through an existing CGM will tend to be denser as it
is more confined (compared to the spherically symmetric case) and because it has the opportunity
to sweep up material. None the less, it is instructive as a point of comparison.

\section{Numerical Methods} \label{sec:methods}
\subsection{Hydrodynamics, gravity and cooling}
We make use of the \textsc{Arepo} code
(\citealt{Springel2010,Pakmor2016} and for the public
release\footnote{\href{https://arepo-code.org}{https://arepo-code.org}} see \citealt{Weinberger2020}).
Gravity is included with a tree-based algorithm.\footnote{A TreePM scheme is available
but not used in this work.} 
\textsc{Arepo} uses a finite volume scheme, solving 
hydrodynamics
on an unstructured, moving mesh. The mesh is defined by the Voronoi tessellation of
mesh-generating points which move with the local fluid velocity (with small corrections
to maintain cell regularity). This gives the scheme quasi-Lagrangian properties,
since cells tend to maintain constant mass over time. However, mass fluxes
between cells are non-zero, so a (de)refinement scheme is typically used to (merge) split
cells to keep them within a factor of two of a desired mass resolution.
Beyond the constant mass (de)refinement scheme, other criteria can be used
to enforce varying mass or spatial resolution within the simulation domain; we make
use of this facility in the schemes described below. We include radiative cooling 
as described in \cite{Vogelsberger2013}. This includes cooling from both primordial
species \citep{Cen1992,Katz1996} and metal lines (in pre-calculated lookup
tables) in the presence of a $z=0$ UV background \citep{Faucher-Giguere2009},
with corrections for self-shielding in dense gas \citep{Rahmati2013}. 
While we do not impose a formal temperature floor, we do not radiatively cool
below $10^4\,\mathrm{K}$.
Finally, we point out that, while omitted here, magnetic fields
can also be included in simulations with \textsc{Arepo} \citep{Pakmor2011} and that
\textsc{Arkenstone} is fully compatible with the
magnetohydrodynamical scheme.

\subsection{ISM effective equation of state and star formation rates}
At the typical resolution at which this model is intended to operate, the multiphase
ISM cannot be well resolved. 
In the present demonstration of the \textsc{Arkenstone} implementation,
we use the model
of \cite{Springel2003} (hereafter \citetalias{Springel2003}),
adopting an effective equation of state (eEoS).
This aims to represent the large scale impacts of small scale
ISM physics (e.g. local stellar feedback, turbulence, molecular cloud
formation and destruction etc.) in an abstract manner, under the assumption that
the ISM reaches an equilibrium configuration such that the pressure
of the unresolved multiphase medium may be determined as a function
of density alone. Following \cite{Vogelsberger2013}, 
to avoid overpressurising the ISM,
we interpolate
between the pressure given by the \citetalias{Springel2003} eEoS 
and an isothermal EoS at
$10^4\,\mathrm{K}$, with a contribution fraction of 0.3 from the 
\citetalias{Springel2003} eEoS.
Gas above a density of $\rho_\mathrm{SF}$ is converted into stars on a
timescale
\begin{equation}
    t_\star = t^\star_0 \left(\frac{\rho}{\rho_\mathrm{SF}}\right)^{-\alpha}.
\end{equation}
We adopt $t^\star_0=2.2\,\mathrm{Gyr}$ and $\rho_\mathrm{SF}/m_\mathrm{p}=0.2\,\mathrm{cm^{-3}}$.
Following \cite{Nelson2019}, we initially use $\alpha=0.5$ (i.e. the
star formation timescale is linearly proportional to the local dynamical time
of the gas), but switch to $\alpha=1$ for gas denser than $228.7\rho_\mathrm{SF}$
(producing more rapid star formation in gas that is dynamically unstable
under the eEoS).

We remark at this point that the \textsc{Arkenstone} model is not dependent
on a specific choice of the ISM EoS, nor on the method used to calculate the
star formation rate (SFR) for a given gas cell. It may therefore be freely used with any
other model \citep[e.g. that proposed by][]{Ostriker2022}, although
it is intended to be used when the ISM cannot be fully resolved.

\subsection{Star and wind particle creation} \label{sec:particle_creation}
We create stellar and wind material from star-forming gas cells
based on their SFR, $\dot{m}_\star=m_\mathrm{cell}/t_\star$. Gas cells produce wind material at a rate of
\begin{equation}
\dot{m}_\mathrm{w} = \eta_M \dot{m}_\star,
\end{equation}
where $\eta_M$ is the input mass loading factor, which we treat as a free
parameter in this work. We stochastically generate star and wind particles
by sampling these production rates, broadly following the method of
\cite{Vogelsberger2013}.\footnote{However, the
wind particle implementation of \textsc{Arkenstone} 
in \textsc{Arepo}
is independent
from that introduced in \cite{Vogelsberger2013}.}

In the case of star particles, we generally convert the
entire cell mass into a star particle. However, we allow for the possibility of
gas cells spawning wind particles of a lower mass. We parameterise the desired
wind particle mass, $m_\mathrm{w}$, relative to the target gas mass resolution of the simulation,
$m_\mathrm{g,tar}$,
via the free parameter: 
\begin{equation}
f_{m,\mathrm{w}} = \frac{m_\mathrm{w}}{m_\mathrm{g,tar}} \leqq 1. \label{eq:f_mw}
\end{equation}
Using wind particles that are of a smaller mass (i.e. higher mass resolution) than
the target gas mass allows for a finer discretisation of the wind injection. This
is particularly important for high specific energy winds which, definitionally,
are characterised by a high ratio of energy to mass loadings.
Insufficient mass resolution will lead to poor
sampling of the wind energy injection rate as highly energetic wind particles are
launched infrequently, creating an artificially bursty behaviour \citep[see e.g.][]{Kim2020a}. 
Additionally,
the use of higher resolution wind particles is an important component of our
hot wind refinement scheme, as detailed in a later section.

Note that we can in principle form more than one wind particle from a single cell in a timestep
if we need to form a large mass of wind material
in a timestep relative to the wind particle mass. In practice, typical timesteps of
gas cells mean that this happens rarely, except for very small values of $f_{m,\mathrm{w}}$.
To avoid leaving arbitrarily low mass gas cells behind, if after spawning wind particles
a cell would be left with less than 10 per cent of the target gas mass resolution, we simply
convert the entire cell. This technically biases the input loadings upwards, but the
effect is negligible.

When a star or wind particle is created, it retains conserved quantities from its parent
gas cell in proportion to its mass (with the exception of internal energy, which is
discarded). Wind particles receive additional ``launch''
energy in both kinetic and thermal forms.
In the full \textsc{Arkenstone} model, the velocity and thermal energy
of the particles are sampled from distributions, following \cite{Kim2020a}.
However, in this work we use a single velocity and
thermal energy in a given simulation, to aid in the clarity
of the demonstration, defined in the following manner.
Wind particles are given a velocity kick with magnitude
\begin{equation} \label{eq:vwind}
\Delta v_\mathrm{w} = \sqrt{\frac{2\eta_{E,\mathrm{kin}}}{\eta_M} u_\star},
\end{equation}
and an internal energy
\begin{equation} \label{eq:uwind}
u_\mathrm{w} = \frac{\eta_{E,\mathrm{th}}}{\eta_M} u_\star.
\end{equation}
$\eta_{E,\mathrm{kin}}$ and $\eta_{E,\mathrm{th}}$ are the kinetic and
thermal input energy loadings, respectively, which we treat as free parameters in this paper.
$u_\star$ is the characteristic specific energy associated with stellar feedback. We adopt $u_\mathrm{\star} = 5.26\times10^5\ (\kms)^2$. This 
corresponds to one supernova of
$10^{51}\,\mathrm{erg}$ for every $95.5\,\mathrm{M_\odot}$ of stellar mass formed, 
which is consistent with the reference value used in \cite{Kim2020a} for a 
\cite{Kroupa2001} initial mass function (IMF).
Of course, a significant quantity of the stellar feedback energy budget is contained
in non-SN channels (e.g. radiation and winds from young stars). However, the purpose
of the $u_\star$ reference value is simply to enable a scaling of available
energy to power galactic winds with SFR, so it is convenient to scale with
the available SN energy (particularly as they are likely the dominant energy
source behind galactic winds).

It can be seen that the
Mach number of the wind at launch is related to the ratio of
the energy loadings as
\begin{equation}
\mathcal{M}_\mathrm{launch} = \sqrt{\frac{2}{\gamma \left(\gamma -  1 \right)}\frac{\eta_{E,\mathrm{kin}}}{\eta_{E,\mathrm{th}}}}.
\end{equation}

The velocity kick can either be applied to the wind particle vertically out of the disc plane\footnote{In the
non-cosmological simulations here, we simply add the kick parallel to the domain z-axis. In a cosmological
simulation, one could, for example, define a direction out of the disc 
by taking the cross product of the gas cell velocity 
vector with the potential gradient in the rest frame of the galaxy \citep{Springel2003}. Alternatively,
one could launch down the local density gradient as an estimate for the path of least resistance.}
or in a random direction (isotropically). It can be seen that
for either approach, momentum is conserved across the ensemble of launched wind particles. 
We adopt the vertical launch direction as our fiducial choice in this work, but also explore the impact of this
choice.
The wind particle
internal energy is not cooled away prior to recoupling.

In this work, for simplicity, the wind particle inherits the metallicity of the gas from which
it was launched. However, our implementation can allow for a relative enrichment of the particle dependent on
its phase \citep[see e.g.][]{Kim2020a}. We will explore this aspect of the model in future work.

\subsection{Wind particle recoupling}
While in-flight, we identify the gas cell that contains the wind particle (the ``host cell'').
Wind particles are initially hydro-decoupled, meaning that they do not interact with ambient gas as they
move except via gravitational forces.
However, once a particle has left the ISM it will recouple back into the local medium. This is
implemented by means of a recoupling density threshold,
$\rho_\mathrm{rec} = f_\mathrm{rec} \rho_\mathrm{SF}$.{}
Our default choice for the value of $f_\mathrm{rec}$ is 0.1, as in \citetalias{Springel2003}.
When the host cell of a wind particle has a density lower than $\rho_\mathrm{rec}$, the particle will recouple
into it.
We have also implemented a maximum travel time, after which wind particles must recouple, to ensure that
particles that fail to escape the ISM and fulfil the density criterion do not exist indefinitely.
However, for all the winds we demonstrate in this work, wind particles never fail to fulfil
the density criterion and typically do not travel more than a few kpc while decoupled 
\citep[a similar situation is reported in a cosmological context in][]{Pillepich2018}. We therefore
disable the travel time criterion in this work.
Recoupling may proceed either by ``standard recoupling'' or ``displacement recoupling'', as described below.

\subsubsection{Standard recoupling}
In a standard recoupling, the wind particle deposits its conserved quantities (i.e. mass, energy, momentum
and metals) into the host cell and is then removed from the simulation. This is equivalent to the approaches
typically adopted in earlier wind particle schemes. However, this form of recoupling will dilute the specific
energy of a wind particle as it is merged with the host cell. This is problematic for
high specific energy winds, especially when combined with high resolution wind particles. A single wind particle
may have a high velocity and temperature, but the total energy increase experienced by a more massive host cell
may be small. In particular, thermal energy can be rapidly radiated away if the resulting temperature increase is
not high enough to clear the peak of the cooling function near $10^5\,\mathrm{K}$. Therefore, in the complete
model, standard recoupling is only our first choice if the host cell is already flagged as a ``hot
wind cell'' (which we define below). Otherwise, we attempt displacement recoupling.

\subsubsection{Displacement recoupling} \label{sec:disp_rec}
If the ISM/CGM interface were properly resolved, high specific energy outflows could
vent out of the ISM through chimneys/superbubbles without mixing into or sweeping up
significant amounts of mass which would drop their temperature and velocity.
Our displacement recoupling scheme is designed to preserve the high specific energy of high resolution (relative to the target gas cell mass) wind particles despite the lack of a porous ISM/CGM transition region.
Ideally, we would like to refine the host cell to a comparable mass resolution to the wind particles.
However, as we
will demonstrate in this work, we typically require refinement by a factor of $10-100$ in mass.
The normal refinement scheme (based on cell splitting) cannot act fast enough for our needs
without producing an extremely irregular mesh structure which breaks the hydrodynamics scheme.
Instead, when a particle triggers the recoupling criterion, we first redistribute material from the host cell to its 
immediate surroundings. We can then fill the host cell with the contents of the wind particle. 
This results in a low mass, high specific energy cell (a hot wind cell). This method has the advantage of maintaining the existing mesh structure. While the energy of a lone hot wind cell would quickly be diluted by the presence of its
neighbours, a population of hot wind cells is rapidly established in the recoupling region as multiple wind particles
recouple in this manner, suppressing numerical overcooling.

When performing a displacement recoupling, we first search for the nearest $N_\mathrm{ngb}$ cells\footnote{More specifically, their mesh generating points.}.
Our default choice is 32 neighbours. Of these, only cells not currently flagged as hot wind cells are eligible
to receive displaced material. If no eligible neighbours are found, we perform
a standard recoupling. We do this, rather than searching within a wider radius, to avoid displacing
material over arbitrarily large distances. This is particularly important when the wind has established itself
because large regions of the ISM/CGM transition region will now be occupied by hot wind cells.
Ideally, we wish to displace all the material in the host cell to the eligible neighbours. However, despite having
a higher specific energy than its neighbouring gas cells, it will almost certainly have a lower density. This is because its volume
is unchanged after recoupling, but we have reduced its mass. Under some circumstances 
(i.e. a very high resolution wind particle combined with low energy loadings), 
this can mean that the host cell
will be underpressured relative to neighbours and the low specific energy, displaced material will shortly flow back in.
We therefore specify a minimum pressure contrast, $\chi_{P,\mathrm{min}}=1.1$, that we wish to achieve after the
displacement coupling is complete. If necessary, we retain a fraction of the original host mass, $f_\mathrm{ret}$,
to ensure the minimum pressure contrast is achieved (at the cost of diluting the specific energy of the wind
particle). 
In \cref{ap:fret} we describe how $f_\mathrm{ret}$ is calculated.
In the simulations presented in this work, it is extremely rare that any mass needs to be retained, but
we include this part of the model as a safeguard.
The mass to be displaced (i.e. the pre-recoupling contents of the host cell, minus any retained fraction)
is re-distributed to the eligible neighbours, weighted by volume and a cubic-spline kernel function. 
Additional conserved quantities of the cell (momentum, energy, metals etc.) are likewise transferred
along with the mass. The conserved contents of the wind particle are then added to the host cell.

In order to identify which material has been injected as a hot wind phase, we use a passive scalar
"dye" which is advected with the gas. When a wind particle recouples into a cell, either by
displacement or a standard recoupling, the mass fraction of the dye in the cell,
$f_\mathrm{w}$, is set to unity
(after any displaced material has been removed).
$f_\mathrm{w}$ will decrease as the hot wind mixes with the undyed gas of the CGM.\footnote{
	Over a long enough time period and given sufficient recycling of wind material, the mass
	fraction of the dye in the ambient CGM will gradually increase. For the simulations
	presented in this work, this effect is negligible. However, in a cosmological simulation
	this may gradually impede the ability to identify hot wind material for the purposes of
	our scheme. Decaying the dye with time solves this problem, but for simplicity we do not
	do this here.
}
A cell is considered flagged as a hot wind
cell if $f_\mathrm{w}$ is above some threshold value, $f_\mathrm{w,thresh}$. 
We take $f_\mathrm{w,thresh}=0.1$
as our fiducial choice, but find our results are relatively insensitive to this value.
As mentioned above, hot wind cells are both ineligible to receive displaced material and will
cause a standard recoupling to be executed if they are the host cell of a recoupling wind particle.
If at any point the temperature of the cell drops below $T_\mathrm{w,thresh}=5\times10^4\,\mathrm{K}$
the mass fraction of the dye is set to zero.

In order to preserve the superior mass resolution of the hot wind cells, flagged cells are subject
to a modified (de)refinement scheme. Without this alteration, the default scheme would immediately
attempt to de-refine the cells back to the target gas mass resolution of the simulation. Additionally,
subsequent recouplings of wind particles into an already refined cell will raise its mass. We therefore
set a new target mass resolution for hot wind cells, $m_\mathrm{wg,tar}$.
The purpose of our refinement
is specifically targeted to correctly resolve the evolution of the wind, \textit{not} to refine the
entire CGM 
(which, while interesting, is considerably more computationally expensive). As we will demonstrate in this
work, the crucial location where enhanced resolution is needed is near the base of the wind. Resolution
can then be coarsened as the wind flows out into the wider halo. We therefore enforce
$m_\mathrm{wg,tar} = m_\mathrm{w}$ (i.e. the initial high mass resolution is preserved)  
inside $0.1r_{200}$, before increasing the target mass linearly as a function radius until
$m_\mathrm{wg,tar} = m_\mathrm{g,tar}$ (i.e. the target gas mass of the whole simulation)
is reached at $0.5r_{200}$. The choice of these radii is
empirical, but results are not sensitive so long as the high resolution is enforced
close to the base of the wind (as we will demonstrate below) 
and the radial evolution is sufficiently gentle that the de-refinement scheme
can operate effectively.

Finally, we note that while this radial resolution dependence is trivial to implement in a non-cosmological
setup (where the galaxy is in the centre of the box), it is more complicated to use in a cosmological
volume or zoom-in. This issue is beyond the scope of this work, but we stress that it is by no means
insurmountable. A radial scale could be enforced with the aid of an on-the-fly group-finder, for example.
Alternatively, an additional passive scalar could be used to impose a time, relative velocity or
relative temperature dependence on the mass resolution, all of which are equivalent to a radial
dependence given a prediction of the emergent wind properties.

\begin{table}
\caption{The parameters defining the two sets of initial conditions
used in this work. For details about the models adopted and the
definition of the symbols, see the
main text. Note that the quantities reported in this table are
all input parameters with the exception of $r_{200}$,
which is derived from the halo mass, concentration and cosmology,
and the CGM mass inside $r_{200}$, which is a derived
quantity of the cooling flow solution given the other constraints.}
\label{tab:IC}
\begin{center}
\begin{tabular}{lrr}
\hline
Parameter & Fiducial ICs & Supersonic ICs\\
\hline
\textbf{Dark matter} & &\\
$M_{200}$ & $10^{11}\,\Msun$ & $10^{11}\,\Msun$\\
$c$         & 10               & 10\\
$r_{200}$ & 97.9 kpc         & 97.9 kpc\\
$s_e$       & 1.5              & 1.5\\
$b_e$       & 1                & 1\\
\hline
\textbf{Stellar disc} & &\\
$M_{\mathrm{disc},\star}$   & $8\times10^{9}\,\Msun$ & $1.6\times10^{9}\,\Msun$\\
$R_\mathrm{s}$ & 2.5 kpc                & 2.5 kpc\\
$z_\mathrm{s}$ & 0.25 kpc                & 0.25 kpc\\
$m_\star$      & $8\times10^{4}\,\Msun$ & $8\times10^{4}\,\Msun$\\
\hline
\textbf{Stellar bulge} & &\\
$M_{\mathrm{bulge},\star}$    & $10^{8}\,\Msun$ & $2\times10^{7}\,\Msun$\\
$r_\mathrm{s}$ & 2.5 kpc                & 2.5 kpc\\
$m_\star$      & $8\times10^{4}\,\Msun$ & $8\times10^{4}\,\Msun$\\
\hline
\textbf{Gas disc} & &\\
$M_{\mathrm{disc},\mathrm{gas}}$    & $2\times10^{9}\,\Msun$ & $4\times10^{8}\,\Msun$\\
$R_\mathrm{s}$ & 2.5 kpc                & 2.5 kpc\\
$T_\mathrm{0}$ & $10^4\,\mathrm{K}$     & $10^4\,\mathrm{K}$\\
$Z_\mathrm{0}$ & $1\,Z_\odot$     & $1\,Z_\odot$\\
$m_\mathrm{g,tar}$ & $8\times10^{4}\,\Msun$ & $8\times10^{4}\,\Msun$\\
\hline
\textbf{CGM}&&\\
$r_\mathrm{circ}$ & 2.5 kpc & 2.5 kpc\\
$r_\mathrm{sonic}$ & 2 kpc & 5 kpc\\
$Z_{0}$ & $0.1\,Z_\odot$ & $0.1\,Z_\odot$\\
$M_\mathrm{CGM}\left(<r_{200}\right)$ & $2.96\times10^9\,\Msun$ & $4.33\times10^9\,\Msun$\\
$m_\mathrm{g,tar}$ & $8\times10^{4}\,\Msun$ & $8\times10^{4}\,\Msun$\\
\hline
\end{tabular}
\end{center}
\end{table}
\begin{table*}
\caption{Parameters governing the different wind models used in this work. 
The first column lists the shorthand name that we use in the text to refer
to the model.
Columns 2-4 give the input mass, kinetic energy and thermal energy loadings.
For reference, the next five columns show quantities derived from the three 
input loading parameters: the total
energy loading, the wind particle launch velocity, the wind particle
temperature, the wind particle sound speed and the Mach number at launch.
The final column denotes whether the new displacement recoupling scheme
(including refinement)
is used instead of standard recoupling.}
\label{tab:wind}
\begin{center}
\begin{tabular}{rrrrrrrrrc}
\hline
Model name & $\eta_M$ & $\eta_{E,\mathrm{kin}}$ & $\eta_{E,\mathrm{th}}$ & $\eta_{E,\mathrm{tot}}$ & $v\,\left[\kms\right]$ & $T\,\left[\mathrm{K}\right]$ & $c_\mathrm{s}\,\left[\kms\right]$ & $\mathcal{M}_\mathrm{launch}$ & Recoupling\\
\hline
\lowspec{} & 6.41 & 0.869 & 0.097 & 0.966 & 378 & $3.79\times10^{5}$ & 94 & 4 &Standard\\
\highspec{} & 0.32 & 0.321 & 0.579 & 0.900 & 1028 & $4.53\times10^{7}$ & 1028 & 1 & Standard\\
\highark{} & 0.32 & 0.321 & 0.579 & 0.900 & 1028 & $4.53\times10^{7}$ & 1028 & 1 & Displacement\\
\medark{} & 0.24 & 0.048 & 0.087 & 0.135 & 456 & $8.91\times10^{6}$ & 456 & 1 & Displacement\\
\hline
\end{tabular}
\end{center}
\end{table*}
\subsection{Initial Conditions}

We simulate idealized, isolated systems comprised of dark matter, a disc and bulge of pre-existing stars, a
gas disc and a CGM/IGM. The input parameters describing our initial conditions (ICs) can be found
in \cref{tab:IC}. In this work, we use lowercase $r$ to denote radii in spherical coordinates and uppercase $R$
for radii in cylindrical coordinates (in the plane of the galactic disc).
The dark matter is modelled as a spherically symmetric, static background potential and includes both an
inner and outer halo component. The inner component follows a Navarro-Frenk-White \citep[NFW,][]{Navarro1997}
profile with $M_{200}=10^{11}\,\Msun$
and a concentration of 10. In combination with a
\cite{PlanckCollaboration2020} cosmology, this gives $r_{200}=97.9\,\mathrm{kpc}$.\footnote{In this work,
$M_{200}$ and $r_{200}$ are defined relative to the critical
density.}
The outer halo component is modelled following \cite{Diemer2014}, with their median values
of $s_e=1.5$ and $b_e=1$. The stellar disc and bulge, and the gas disc are generated using
the code \textsc{MakeNewDisk} \citep{Springel2005}. The stellar and gas discs have exponential
surface density profiles with a scale length of $R_\mathrm{s}=2.5\,\mathrm{kpc}$. The stellar disc has a gaussian vertical
density profile with a scale height of $z_\mathrm{s}=0.25\,\mathrm{kpc}$. The gas disc has a vertical density profile that
is set to produce hydrostatic equilibrium at its initial temperature of $T_0=10^4\,\mathrm{K}$. 
We truncate the gas disc beyond five scale lengths and five scale heights.
The disc has an initial metallicity of $Z_{0}=1\,Z_\odot$ (where we adopt $Z_\odot=0.0127$).
The stellar bulge is spherically symmetric and follows a \cite{Hernquist1990} density profile
with a scale length of $r_\mathrm{s}=0.25\,\mathrm{kpc}$. Our fiducial set of initial conditions uses
$M_{\mathrm{disc},\star}=8\times 10^9\,\Msun$, $M_{\mathrm{bulge},\star}=10^8\,\Msun$ and $M_\mathrm{disc,gas}=2\times 10^9\,\Msun$,
providing a strong engine to drive
powerful winds, which is useful for our demonstrations.
We also perform a simulation that has a factor five lower mass in these three
components in \cref{sec:supersonicICs} (the ``Supersonic ICs'' in \cref{tab:IC}).

We initialize the CGM gas in a steady state rotating cooling flow configuration, the full details of which can be found in \cite{Stern2023}. 
This solution corresponds to the expected CGM solution without ongoing heating by feedback (but accounts for enrichment/depletion of the CGM by previous feedback events) and hence is a reasonable solution to use in a simulation which independently implements feedback such as ours.
The density, temperature, and radial velocity profiles of the CGM gas are first set according to a non-rotating cooling flow solution in which there is a constant radial mass flux and the radiative cooling time approximately balances the compressive heating time \citep{Stern2019}. The CGM gas is then given an azimuthal velocity $v_\phi = v_{\rm circ} \sin \theta (r/r_{\rm circ})^{-1}$, where $v_{\rm circ}$ is the circular velocity of the potential and $r_{\rm circ}$ is the circularization radius.
We remove CGM gas that falls inside the volume of our gas disc, which ensures $v_\phi$ does not diverge at small $r$. This also means
that the transition between the disc and the CGM is initially discontinuous.
This results in a short-lived disruption of the very edge of the disc,
but no long-lasting impact on the evolution of the system.
It is important that the gas distribution extends sufficiently far that there is a large enough mass reservoir to sustain the cooling flow and that the region of interest (i.e. the halo) is isolated from unphysical behaviour arising from the boundary conditions over the timespan of the simulation.
For these reasons, our CGM/IGM gas component initially
extends to $600\,\mathrm{kpc}\sim6r_{200}$.
However, we do not require our full resolution far outside the halo. Therefore,
beyond $200\,\mathrm{kpc}$ the gas mass resolution smoothly coarsens as a function
of radius such that it increases by a factor of three every
$\sqrt{2}\times200\,\mathrm{kpc}$. This radial dependency is also enforced
during the simulation by appropriate modifications to the (de)refinement scheme.
Outside of $600\,\mathrm{kpc}$, we fill the remainder of the
$\left(2.4\,\mathrm{Mpc}\right)^3$ volume with a coarse grid of ``vacuum'' cells.

The nature of cooling flows is such that they must always undergo a transition from subsonic radial motion at large radii to supersonic radial motion at some finite sonic radius, $r_{\rm sonic}$. With all else being equal cooling flows with larger mass flux have larger $r_{\rm sonic}$. 
A sonic transition can however be avoided if $r_{\rm circ}>r_{\rm sonic}$, in which case the subsonic flow would spin up and cool into the disc rather than turn into a supersonic flow.
Recently it has been shown in both idealized and cosmological simulations that many CGM properties and the nature of how galactic feedback couples to the CGM change dramatically when $r_{\rm sonic}$ is greater than or less than $r_{\rm circ}$ \citep{Stern2020,Stern2021}. In our fiducial simulations we adopt a cooling flow with $r_{\rm sonic} < r_{\rm circ}$ (i.e., in the subsonic limit), but in \cref{sec:supersonicICs} we compare against simulations with CGM initial conditions such that $r_{\rm sonic} > r_{\rm circ}$ (i.e., in the supersonic limit). The parameters for these
flows can be found in \cref{tab:IC}.

Finally, after generating the initial distribution of mesh-generating points
via Poisson sampling, we perform the mesh relaxation procedure
described in \cite{Springel2010}, section~4.3, to produce a well-structured
mesh with regular, round cells. We use a target gas mass of 
$m_\mathrm{g,tar}=8\times10^4\,\Msun$ (but present coarser
resolution tests in \cref{ap:base_res}). Recall, however, that gas cells can have a smaller
mass if they are subject to the \textsc{Arkenstone} hot wind cell refinement
scheme or a larger mass if they are far outside the halo (as described above).
The stellar particles initially present in the simulation similarly
have a mass of $m_\star = 8\times10^4\,\Msun$. The gravitational
softening length of collisionless particles in the simulation
is fixed at 195~pc. For gas cells, we employ adaptive softening
lengths: the softening is 2.5 times the cell radius, down to a minimum
softening of 50~pc.

\subsection{Wind model parameters}
\Cref{tab:wind} lists the input loading factors for the wind models used in this work, along with some derived
quantities (velocity, temperature and sound speed) for reference. It also shows whether standard
recoupling is used or the new displacement recoupling and refinement scheme. The \lowspec{}
model uses loading factors that are representative of those that would be typically used for a
galaxy of this mass in existing cosmological simulations (see the introduction of this work
for references).
It has a near unity energy loading, but is significantly mass loaded: 90\% of the
energy is in the kinetic component.
In the \highspec{} model
we use almost the same total energy loading but a factor twenty lower mass loading. By coupling the same
amount of energy to a much smaller mass, we achieve a very high specific energy wind. We also re-balance the
kinetic and thermal energy loadings such that $\mathcal{M}_\mathrm{launch}=1$. \highark{} has the same
loadings as \highspec{} but uses the new displacement recoupling and refinement scheme. 
These three models
are used in our fiducial simulations. 
We do not test the new scheme with the \lowspec{} loadings because a low specific energy wind such
as this does not suffer from the numerical issues described in \cref{sec:challenge}.
We use the \medark{} parameters in \cref{sec:twindfix}; more details
can be found there.

\section{Results} \label{sec:results}
\subsection{Fiducial Simulations} \label{sec:fid_sims}
\subsubsection{Outflow morphology}

\begin{figure*} 
\centering
\includegraphics{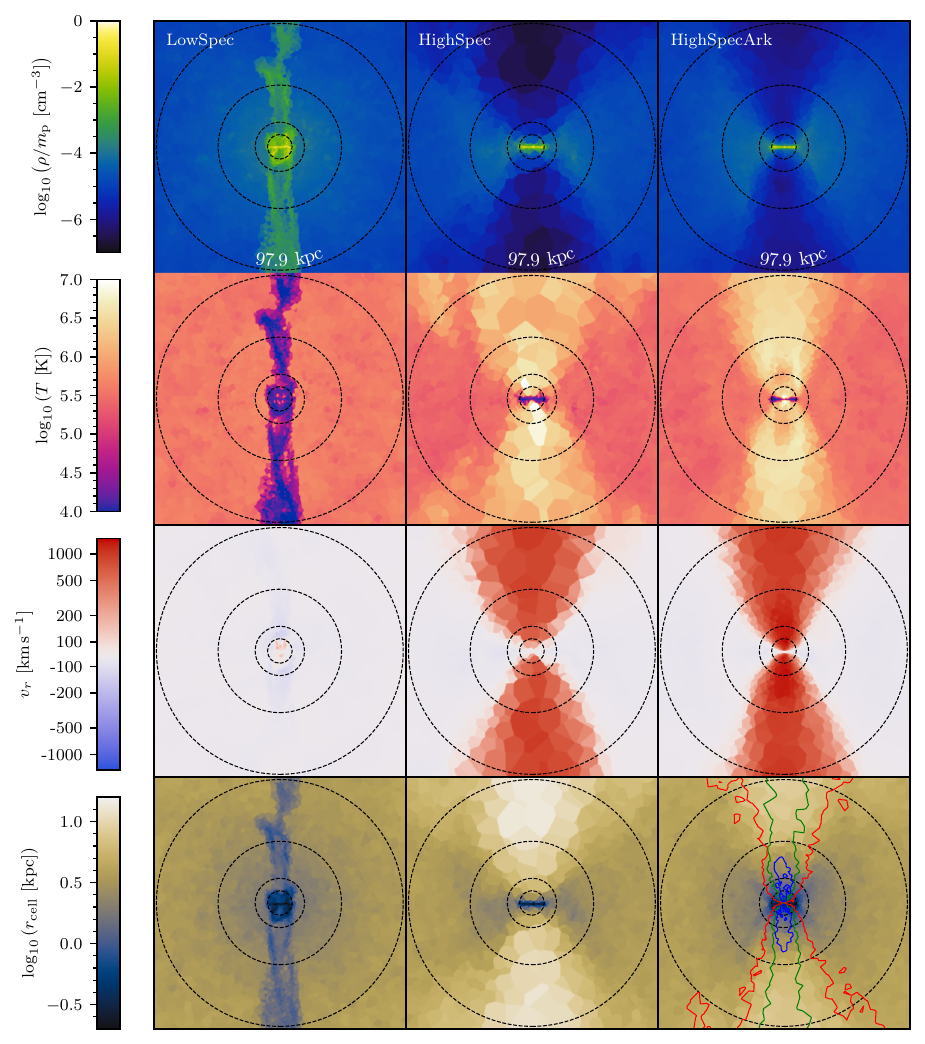}
\caption{Slices through the domain for the \subfive{} simulations at 1.5 Gyr.
The slices cut vertically through the galaxy and the wind. From left to right,
we show the simulations with the \lowspec{}, \highspec{} and \highark{} models.
From top to bottom, we show gas density, temperature, radial velocity
and cell radius. The dashed circles indicate 0.1, 0.2, 0.5 and 1~$r_{200}$.
For the \highark{} cell radius slice, we overlay contours corresponding
to the mass fraction of the hot wind dye. The red, green and blue contours
correspond to mass fractions of 0.1, 0.5 and 0.75, respectively.
The \lowspec{} wind produces a narrow column of cold, dense gas which is
already infalling by 1.5~Gyr, while a low altitude fountain is present.
The \highspec{} and \highark{} winds are fast, hot and of lower density
than the surrounding CGM. They fill a larger volume than the \lowspec{}
wind. The higher spatial resolution provided by the refinement scheme used
in \highark{} is evident.}
\label{fig:slice_subsonic_5mdisk} 
\end{figure*}

In \cref{fig:slice_subsonic_5mdisk} we show images of the
\subfive{} simulations with \lowspec{}, \highspec{} and \highark{} winds
at 1.5~Gyr.
These are slices through the centre of the domain 
oriented along the axis of rotation of the galaxy/CGM 
and therefore the wind launch direction. We show density,
temperature, radial velocity and cell radius. For the latter
quantity, this is determined from the cell volume approximating
the cell as a sphere. 
We also overlay circles at 0.1, 0.2, 0.5 and 1~$r_{200}$.
These reference radii will be used in our later analysis.

These images are instructive in that,
at a glance, 
they show the striking contrast in how low and high specific
energy winds evolve.
For the \lowspec{} simulation, the wind is visible as a column of
gas which is significantly overdense and cold with respect to
the ambient CGM. 
This material
is highly collimated at all radii, showing the imprint of
the choice of vertical launch direction.
The vast majority of the CGM is unaffected by the wind.
By this point in the simulation, mass from the
ISM has been thrown as far as $r_{200}$ and beyond.
The most far flung material originates from an initial burst
of star formation at the beginning of the simulation.
However,
the outflow has stalled, with a significant fraction of the
wind beginning to flow back down towards the galaxy.
The picture is more complex close to the centre of the system,
where a low altitude fountain flow is set up. This leads to a
significant churning of material within a few tens of kpc of
the disc plane.

The higher specific energy winds, \highspec{} and \highark{},
have broadly similar features to each other. Unlike the overdense wind
produced in the \lowspec{} simulation, they carve out
a significantly underdense region compared to the CGM.
Likewise, the wind is much hotter than the ambient medium.
The outflows occupy a significantly larger volume, spreading
to achieve a greater opening angle. 
All of the material
within the wind is flowing out of the system at high velocity,
with no recirculating fountain flow present. 

While the general
morphologies of both high specific energy simulations are similar,
it can be seen that the full \textsc{Arkenstone} model has a major effect
on the sizes of cells (by design). This is apparent both in the
maps of cell radius directly, but also in the outlines of cells
visible in the slices.\footnote{Note that in these images we
show the properties of the cell within which the pixel falls
without taking into account gradients within the cell.} 
Without \textsc{Arkenstone}'s displacement recoupling
and refinement scheme, cells in the wind in the \highspec{} simulation
are highly spatially extended due to the low densities, often
having sizes comparable to their distance from the galaxy.
The poor spatial resolution means that the properties of the gas on
the ISM-wind interface are very noisy. When viewing the time
evolution of these maps, the wind can be seen to be driven in
an extremely bursty manner, featuring rare events in which individual
cells are heated and inflated upwards away from the disc. In contrast,
the lower right panel of \cref{fig:slice_subsonic_5mdisk} shows
the refinement scheme at work in the \highark{} simulation. Within
0.1~$r_{200}$ the spatial resolution remains high despite the low
densities. Individual cells begin to become visible in the maps at larger
radii as we relax the refinement constraints and the density drops.
The structure of the wind is much smoother close to the galaxy and the
burstiness (when viewed in a time series) is essentially removed. This is
partly because of the higher spatial resolution, but also due to the
better energy resolution of the wind particles (as discussed earlier).
We will examine these differences more quantitatively later in this work.

\begin{figure} 
\centering
\includegraphics{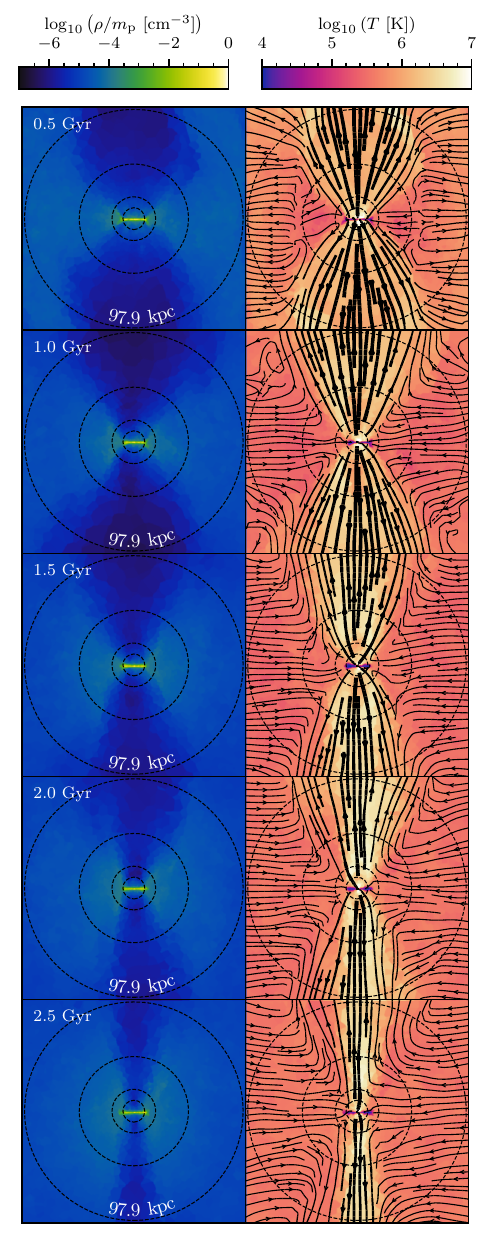}
\caption{Slices showing density (left) and temperature (right),
as in \cref{fig:slice_subsonic_5mdisk},
for the \subfive{} \highark{} simulation, showing how the
opening angle of the outflow reduces as a function of time.
Additionally, we overplot streamlines on the temperature map
with line thickness indicating the magnitude of the velocity
on a linear scale. The wind bicone narrows as the
simulation progresses.}
\label{fig:slice_dens_subsonic_5mdisk_highspec_ark} 
\end{figure}

In \cref{fig:slice_dens_subsonic_5mdisk_highspec_ark} we show a time-series
of density and temperature slices for the \highark{} simulation.
We also overplot velocity streamlines on the temperature map.
It can be seen
that the shape and opening angle of the outflow change
significantly over time. At early times, the outflow rapidly expands
to displace a significant region of inflowing CGM. 
In addition to the gas moving rapidly along the axis of the wind
it can be seen in the 0.5~Gyr image that there is a general
expansion of the CGM with inflows only occurring within
0.5~$r_{200}$ in the disc plane.
However, as the
system evolves, the wind is restricted to an increasingly
narrow region and by 2.5~Gyr has approximately the same width 
as the disc. The behaviour
is similar for the \highspec{} simulation (the \lowspec{} simulation
maintains the slender column morphology shown in 
\cref{fig:slice_subsonic_5mdisk} at all times). The change
in the wind morphology is driven by the gradual reduction of
injected power over time (caused by a declining SFR, see the next section
for details), limiting the ability to carve out such a large cavity
in the inflowing CGM. The transition in the wind geometry leads to a modification
of the radial evolution of the flow which we will explore later in this
work.

\subsubsection{Mass and energy flows} \label{sec:flows}
\begin{figure*} 
\centering
\includegraphics{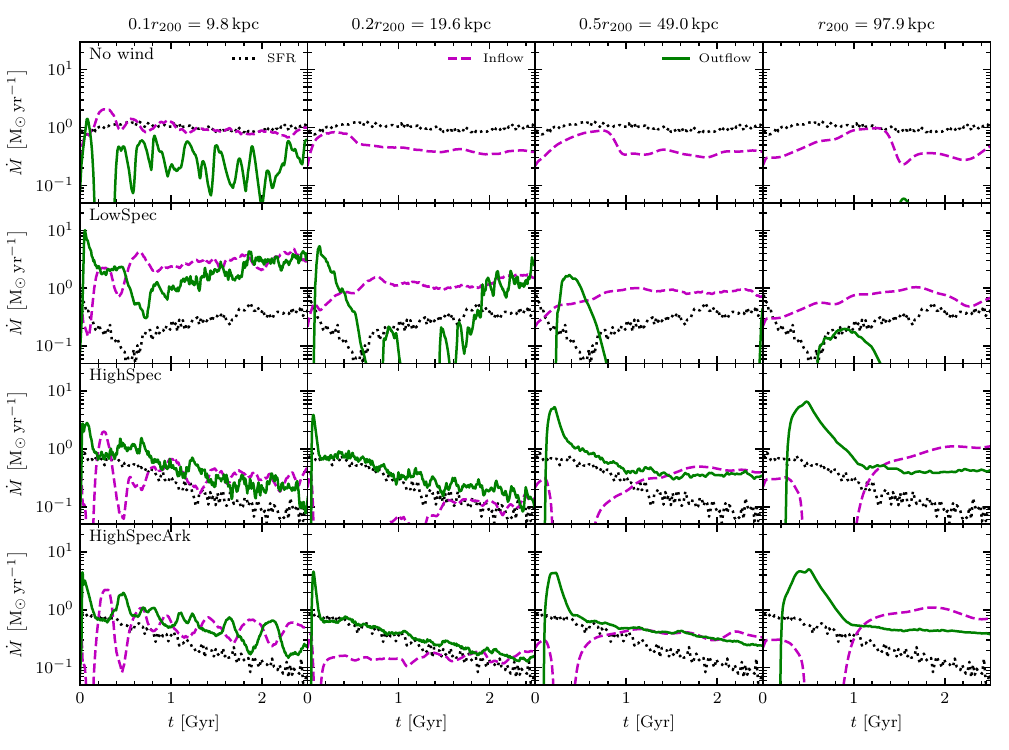}
\caption{SFR, mass inflow and outflow rates through spheres
at various radii for the \subfive{} simulations. Rows from top to bottom
show the no wind, \lowspec{}, \highspec{} and \highark{} simulations,
respectively.
Columns from left to right show measurements at 0.1, 0.2, 0.5 and 
1~$r_{200}$, respectively. For a given simulation, we repeat the
SFR in each column for reference. The \lowspec{} simulation
reduces the SFR significantly relative to the no wind case by
ejecting large quantities of the ISM. This material is kept circulating
in a low altitude fountain flow. By contrast, the \highspec{} and \highark{}
simulations are far more efficient at suppressing inflows from the CGM. This
gradually leads to a reduction in SFR as the supply of star forming gas is
used up without replenishment.}
\label{fig:mflows_subsonic_5mdisk} 
\end{figure*}
We now turn our attention to the flow of mass and energy in the system.
In what follows, we calculate the SFR of the galaxy,  $\dot{M}_\star$,
at a given time from the mass of stars formed in the preceding 10~Myr.
Defining the mass and energy fluxes through a region or surface is
more complicated. When using Lagrangian codes, inflows and outflows
are often measured by searching for resolution elements (in the case of
\textsc{Arepo}, usually the mesh generating points or the cell 
centers of mass) within a slab or thick shell 
\cite[for one example of the procedure, see][section 5.4]{Smith2021a}.
However, given the huge dynamic range in spatial resolution both within
and between different simulations, this approach
is unsuitable for this work. Instead, we define our fluxes with reference
to a thin sphere at a radius, $R$, centred on the
galaxy centre.
We discretise the sphere
into $N_\mathrm{pix}$ equal area pixels using the \textsc{HealPix} library
\citep{Gorski2011}. We use $N_\mathrm{pix}=786432$ such that the inter-pixel spacing is finer
by a factor of a few than the diameter of the smallest cells intercepted
by our reference spheres. For each pixel, we find the gas cell within
which it is located and map the gas cell's properties
onto the
pixel.\footnote{We do not attempt to reconstruct the gradients within
the cell and extrapolate to the pixel location.} The mass and energy
fluxes per unit area through a pixel are then
\begin{equation}
\mathcal{F}_M = \rho v_r,
\end{equation}
\begin{equation}
\mathcal{F}_E = \rho v_r \left(\frac{1}{2} v^2 + \frac{1}{\gamma - 1} c_\mathrm{s}^2 \right), \label{eq:fE}
\end{equation}
respectively, for cell density, $\rho$, radial velocity (i.e. normal to the sphere), $v_r$,
magnitude of the total velocity, $v$, sound speed $c_\mathrm{s}=\sqrt{\gamma P / \rho}$ and
$\gamma = 5/3$.
We can also split the energy flux into kinetic and thermal components by considering
only the first or second terms, respectively, inside the parentheses in \cref{eq:fE}.

\begin{figure*} 
\centering
\includegraphics{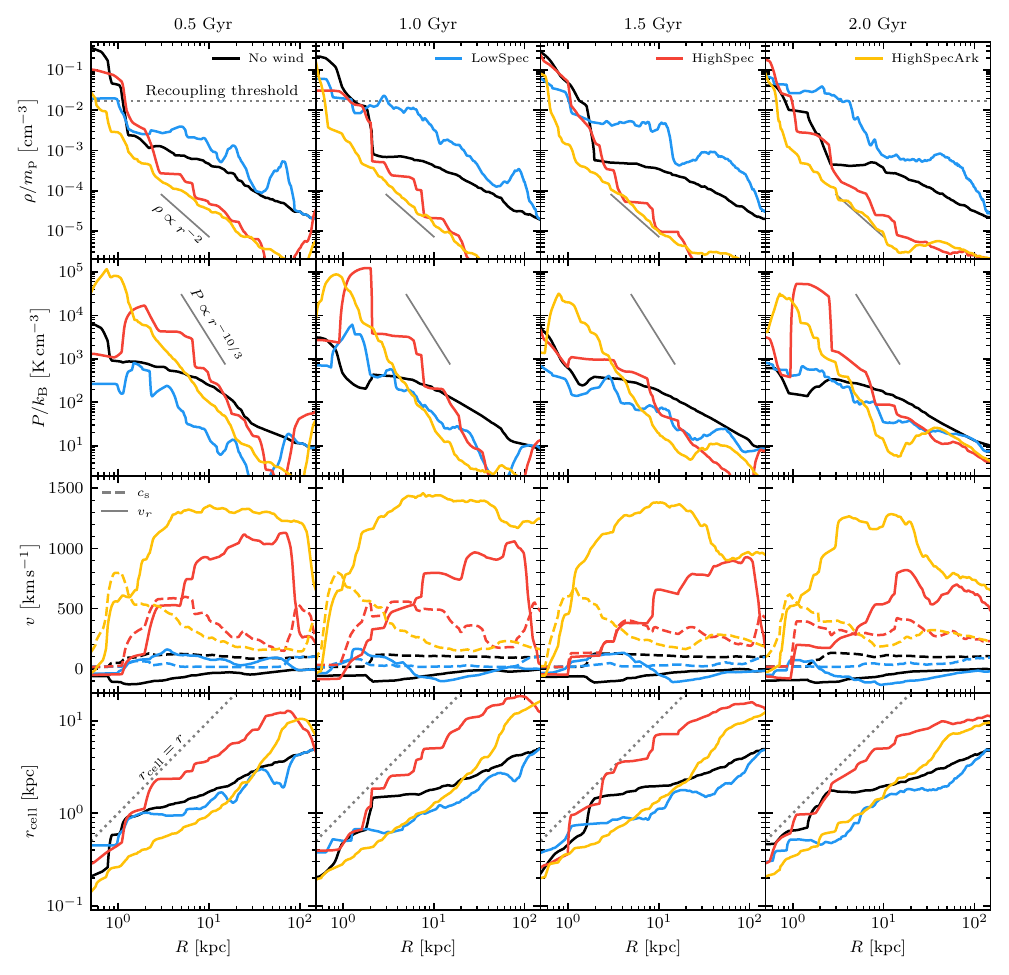}
\caption{For the fiducial simulations,
profiles of various quantities as a function of galactocentric radius
measured within a bicone of opening angle $10\degree$ orientated along
the rotational axis of the disc/CGM. Different columns show different
times throughout the simulation. In the top row, we plot density.
The horizontal grey dotted line indicates the wind particle recoupling threshold 
density. For reference, we indicate the $\rho \propto r^{-2}$ slope of the
asymptotic
\citetalias{Chevalier1985} solution (with arbitrary normalisation).
The second row shows pressure, with an indication of the $P \propto r^{-10/3}$
slope of \citetalias{Chevalier1985}.
The third row shows sound speed (dashed line) and radial
velocity (solid line).
The bottom row shows the cell radius.
In the \highark{} simulation, the wind rapidly accelerates as it flows outwards,
passing through a sonic point within $\sim 1\,\mathrm{kpc}$. The \highspec{} simulation
has much coarser spatial resolution. The acceleration is much more gradual and the
sonic point does not occur until the galactocentric radius is resolved by several cells,
which occurs much further out.
}
\label{fig:profiles_subsonic_5mdisk} 
\end{figure*}

At this point we select only pixels with positive (negative) $v_r$ in order to compute
the outflow (inflow) rates as:
\begin{equation}
\dot{M}_\mathrm{out(in)} = A\sum \mathcal{F}_M,
\end{equation}
\begin{equation}
\dot{E}_\mathrm{out(in)} = A\sum \mathcal{F}_E,
\end{equation}
where the sum runs over the selected pixels, each with equal area $A = 4\pi r^2/N_\mathrm{pix}$
(note $N_\mathrm{pix}$ is the total number of pixels on the sphere, not the number of pixels
in the outflow or inflow selection). Emergent outflow loading factors are then defined in a similar
manner to the input loading factors described in \cref{sec:particle_creation},
\begin{equation}
\eta_M = \frac{\dot{M}_\mathrm{out}}{\dot{M}_\star},
\end{equation}
\begin{equation}
\eta_E = \frac{\dot{E}_\mathrm{out}}{u_\star \dot{M}_\star}.
\end{equation}
For simplicity, we make an instantaneous comparison between the flow rate and the SFR 
(averaged over the preceding 20 Myr) without attempting to correct for travel time 
of the outflow from the ISM to the reference sphere.
We use
reference spheres with radii 0.1, 0.2, 0.5 and 1 $r_{200}$.
For reference,
travel time to $r_{200}$ at 100~$\mathrm{km\,s^{-1}}$ is $\sim1$\,Gyr,
with outflow velocities in this work ranging from $\sim50-1500\,\mathrm{km\,s^{-1}}$
(see \cref{fig:profiles_subsonic_5mdisk}).

In \cref{fig:mflows_subsonic_5mdisk} we show mass inflow and outflow
rates, along with the SFR, for the \subfive{} simulations with
the \lowspec{}, \highspec{} and \highark{} winds, as well as a simulation
without a wind. In the absence of a wind, the SFR remains at a
relatively constant value of $~1\,\Msun\,\mathrm{yr}^{-1}$. This
is approximately equal to the inflow rate through $0.1~r_{200}$,
showing that essentially all CGM material that makes it to the
centre of the halo is converted to stars. There is a small
outflow component through $0.1~r_{200}$, but this is composed
of material at the edge of the disc crossing the reference
sphere. All material at 0.2, 0.5 and 1 $r_{200}$ is inflowing.
Slight instabilities caused by our idealized setup mean the
inflow rates are not completely constant over the 2.5~Gyr shown
(traces of a low amplitude, outgoing soundwave can be detected in the
inflow rates), but this is a minor effect.

The \lowspec{} simulation rapidly suppresses star formation relative
to the no wind case. This is achieved by ejecting large
quantities of the star forming ISM, a consequence of the
significantly super-unity input mass loading factor of 6.41.
The material of this opening burst 
can be seen flowing outwards through $0.1~r_{200}$ immediately
before arriving at larger radii at later times. The final trace of
this initial outflow crosses $r_{200}$ between 0.5 - 1.5~Gyr,
albeit reduced by more than an order of magnitude. The plume
seen in \cref{fig:slice_subsonic_5mdisk} is composed of
this material. The sudden drop in SFR after this first
ISM ejection event naturally leads to a reduction in the
outflow rate by a factor of $\sim30$ within the first 0.7~Gyr.
This allows the inflow rates through $0.1~r_{200}$, initially 
suppressed, to recover to levels comparable to the no wind 
simulation. This in turn leads to a gradual rise in the SFR
as the supply of gas in the disc is replenished. The outflow
rates through $0.1~r_{200}$ continue to track the SFR closely
until inflow and outflow rates are in rough equivalence from
$\sim1.8$~Gyr onwards. The inflow rate is a factor of 2-3
higher than the no wind case. This is a signature of the
significantly mass loaded, low altitude galactic fountain
that this wind model establishes. In the no wind case,
inflowing CGM material crosses $0.1~r_{200}$ once before
joining the ISM and being converted to stars.
However, 
with the \lowspec{} wind active, a unit of ISM mass is
$\eta_M = 6.41$ times more likely to be ejected than 
to turn into a star. This leads to gas cycling in and out
of the ISM multiple times,
enhancing the inflow rates in the inner halo. Inflow rates
remain slightly enhanced relative to the no wind simulation
even out as far as $0.5~r_{200}$, although this extra
material is predominantly associated with the initial
burst. After 2~Gyr, outflow rates balance inflow rates
through $0.2~r_{200}$, but no longer range outflow
is established again during the 2.5~Gyr simulated.

The \highspec{} and \highark{} simulations have very similar
mass fluxes, as shown in \cref{fig:mflows_subsonic_5mdisk}.
The SFR starts at the same level as the preceding simulations.
Unlike the \lowspec{} case, given the sub-unity mass loading 
factor of 0.32, these
winds are unable to suddenly reduce the SFR by ejecting ISM
material. Definitionally, more ISM gas is consumed by star formation
than by conversion to wind material. However, the high specific
energy winds are able to efficiently suppress inflow rates from the
CGM at all radii. This preventative feedback cuts off the
supply of new gas to the ISM, leading to a steady decline in the
SFR.
Despite the sub-unity input mass loading factor, the
outflow rates remain at or above the magnitude of the SFR
at all radii. This is indicative of material being swept
up in the outflows. Close to the centre, at $0.1~r_{200}$
we see a slight oscillatory pattern to the inflows and
outflows, with opposite phase. This is clearer in the
\highark{} simulation which has a generally smoother 
time evolution of the flow rates.
This feature is not transferred to the SFR, which is
isolated from rapid fluctuations of the inflow rates
since it is regulated on the consumption timescale of the
ISM. By 2.5~Gyr, the falling SFR is a factor
$\sim4-5$ lower than the \lowspec{} case (which is rising).
Inflow and outflow rates at small radii 
are also around an order of magnitude lower, but 
those outflows make it all the way out of the halo. 
At larger radii, particularly on the boundary of the
halo, the shape of the time evolution of the inflow and
outflow rates are essentially unrelated. The reason for this
can be readily seen in the images in 
\cref{fig:slice_subsonic_5mdisk,fig:slice_dens_subsonic_5mdisk_highspec_ark}.
There are distinct inflow and outflow regions
i.e. material near the axis of the wind is always
outflowing, whereas the rest of the CGM is always inflowing.
This is unlike the
\lowspec{} simulation which is characterised by a co-spatial
churning of material, with inflows falling back down the path
from which an outflow originated.
In the high specific energy wind simulations,
as the region occupied by the wind reduces in angular
extent at late times
(see \cref{fig:slice_dens_subsonic_5mdisk_highspec_ark}),
the inflow rates at large radii rise towards similar levels to
the no wind simulation. It is possible, therefore, that if
we ran the simulation for a longer period that this would eventually
lead to a replenishment of the ISM as this material
made its way to the centre, a rise in SFR and a corresponding
increase in wind power. However, without a full cosmological
context, extending the simulation into this regime
would provide little physical insight into the
evolution of real galaxies.

\begin{figure} 
\centering
\includegraphics{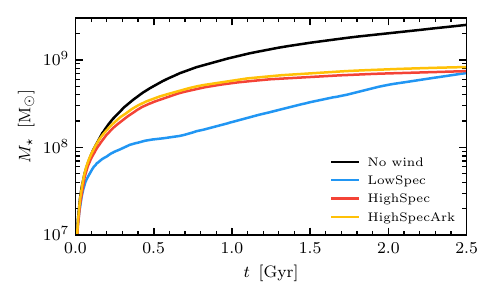}
\caption{The cumulative mass of newly formed stars as a function
of time for the \subfive{} simulations with no wind and \lowspec{}, \highspec{}
and \highark{} winds. Despite regulating the SFR by different methods,
ejective vs. preventative feedback, the low and high specific energy
winds form the same mass of stars by 2.5~Gyr. However, the \lowspec{}
simulation has a higher SFR by this point so is about to overtake
the high specific energy simulations.}
\label{fig:cumul_star} 
\end{figure}
\begin{figure*} 
\centering
\includegraphics{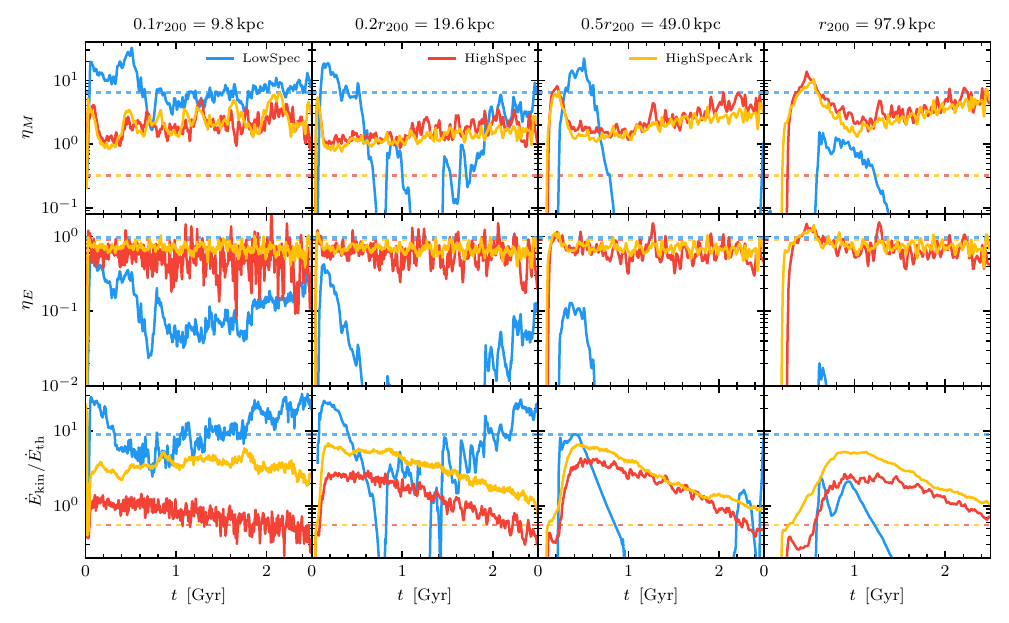}
\caption{Mass loadings (top row), energy loadings (middle row)
and ratio of kinetic to thermal energy fluxes (bottom row)
for the \subfive{} simulations with \lowspec{}, \highspec{}
and \highark{} winds. Columns from left to right show 
measurements at 0.1, 0.2, 0.5 and 
1~$r_{200}$, respectively. 
Dashed lines indicate the input loadings or ratio.
The high mass loading of the
\lowspec{} wind is not maintained far from the galaxy, with
most of the material cycling back in a fountain flow.
The \highspec{} and \highark{} winds increase the mass loading
factor compared to the input value, indicating that the wind is
sweeping up ambient material. They maintain their input energy
loading all the way to the edge of the halo. However, despite
having similar net mass and energy loadings, the \highark{}
wind experiences a rapid increase in the ratio of kinetic to
thermal energy, which is not seen in the \highspec{} simulation
until larger radii.}
\label{fig:load_relative_subsonic_5mdisk} 
\end{figure*}

In \cref{fig:cumul_star} we plot the cumulative mass of newly formed stars
as a function of time (i.e. the integral of the SFR shown in 
\cref{fig:mflows_subsonic_5mdisk}). This highlights the different
manner in which the low and high specific energy winds regulate star
formation. In the no wind case, the SFR remains relatively constant in
time as the ISM is replenished with inflowing CGM. Correspondingly,
the stellar mass builds up linearly with time (note \cref{fig:cumul_star}
is a semi-log plot). The \lowspec{} simulation quickly reduces the
SFR by ejecting a significant fraction of the ISM. When its SFR is at
its lowest, around 0.5~Gyr, it has formed a factor of $\sim5$ less
stellar mass. However, the SFR then rises gradually throughout the
rest of the simulation leading to a marginally super-linear growth
of stellar mass. The \highspec{} and \highark{} simulations are nearly
identical. Unlike the \lowspec{} simulation, they do not rapidly
arrest star formation and so initially track the no-wind case, 
building up stellar mass early.
However, the SFR drops throughout the simulation as the ISM
is consumed and only partially replenished, so the growth of further
stellar mass is curtailed. The result is that \lowspec{},
\highspec{} and \highark{} produce the
same stellar mass by the end of the simulation at 2.5~Gyr.
At this point, the SFR is higher for the \lowspec{} simulation
so it is about to overtake in stellar mass. On the other hand,
as we showed in \cref{fig:mflows_subsonic_5mdisk}, there appears
to be an increased inflow rate making its way inwards in the
\highspec{} and \highark{} simulations which may increase the
SFR in the future.
Nonetheless, as we remarked above, making predictions about the future
state of this system is not particularly useful given the lack
of cosmological context.

\cref{fig:load_relative_subsonic_5mdisk} shows the 
measured outflow mass loadings, energy loadings and the
ratio of the kinetic to thermal components of the 
outflowing energy flux for the same simulations
(omitting the no wind case, which has no significant
outflows) and measurement radii as 
\cref{fig:mflows_subsonic_5mdisk}.
We also mark the input loadings and initial kinetic to
thermal ratio at injection with horizontal dashed lines.
At $0.1~r_{200}$, the \lowspec{} wind maintains its
input mass loading. In other words, essentially all
of the star forming material ejected from the disc,
but no more\footnote{The slight excess mass loading
in the first 0.6 Gyr is largely a consequence of the
travel time induced lag between the rapidly dropping
SFR and outflow rate. The higher specific energy wind
simulations are less affected by the lag because
they are much faster and their SFR does not change
as rapidly.}, makes it to that altitude. However,
because most of this material turns around at larger radii
(as seen in \cref{fig:mflows_subsonic_5mdisk}),
mass loadings are much lower than the input value
further away from the system centre. By contrast,
the two high specific energy wind simulations
demonstrate mass loading factors at all radii that
are almost an order of magnitude larger than their
input value such that they end up with a slightly
super-unity mass loading. This indicates that
the wind is sweeping up additional material on top
of what was ejected from the ISM in the form of
wind particles. Unlike the \lowspec{} model, these
super-unity loadings are maintained all the way
to $r_{200}$. The \highspec{} and \highark{}
models show very similar results, albeit with
a marginally noisier time evolution without the
full \textsc{Arkenstone} model.

We see that the energy loadings of the
\lowspec{} wind are reduced by approximately a factor
of 10 from its input loadings by $0.1\,r_{200}$.
The wind is slowed by gravity and its passage through
the surrounding CGM, draining its energy content. This
means that its initially high mass loadings are
unsustainable as a function of radius as the wind
coasts to a halt. By contrast, the high specific energy
winds maintain their input energy loading almost
exactly at all radii. Despite injecting essentially
the same energy per unit star formation as the \lowspec{}
simulation (they have almost identical input energy 
loadings), the high specific energy of these winds mean that
they are much less susceptible to energy loss. The higher
velocity means that gravitational deceleration is
less important, in a relative sense.
Additionally, radiative cooling losses as the wind flows outwards are
negligible because they have a much lower density and
the higher temperature
moves them away from the peak of the cooling function.
While the broad behaviour is the same for
both high specific energy simulations, we see that
the energy loadings are significantly burstier for the
\highspec{} simulation than the \highark{} run, particularly
at small radii. This burstiness is a direct consequence
of the poor energy resolution of the \highspec{} scheme.
Our improved \textsc{Arkenstone} model smooths out the energy
injection with the aid of lower mass wind particles,
but avoids incurring spurious overcooling because of our
new displacement recoupling scheme, such that the
overall energy fluxes are consistent with the \highspec{}
simulation.

Examining the ratio of the kinetic to thermal components of
the energy fluxes, we see that the \lowspec{} simulation
largely maintains the partitioning imposed by the initial
loading factors ($90\%$ in kinetic) at 0.1~$r_{200}$.
At larger radii, the ratio drops through the early stages
of the simulation; this corresponds to gradual stalling
of the initial outflow.
At later times, at 0.2~$r_{200}$
there is a slight rebalancing in favour of kinetic
energy. This implies that radiative losses are marginally
higher than those incurred via deceleration. This is
consistent with the increasing wind density as a function
of time that we will show below. At 0.5 and 1~$r_{200}$,
the wind
essentially drifts slowly over the reference sphere for
much of the simulation, so the ratio tips in favour of the
thermal component. However, given that the net energy flux
is extremely low, this is of little importance.

Turning to the high specific energy winds,
the ratio of kinetic to thermal energy fluxes reveals a
major difference between the evolution of winds driven
by a more classical wind scheme and our new
model. With the full scheme switched on in \highark{},
we see that prior to reaching $0.1~r_{200}$ the energy
components are rebalanced from the injected 
$\dot{E}_\mathrm{kin}/\dot{E}_\mathrm{th} = 5/9$
(which corresponds to $\mathcal{M}=1$)
to a factor of $\sim2-5$ more kinetic than
thermal content.
This is an expected behaviour:
for an adiabatic wind in an idealised spherical
geometry, \citetalias{Chevalier1985} show
that immediately outside of the injection/star forming region the wind 
undergoes a sonic transition in which the velocity rapidly increases 
then asymptotes while the sound speed rapidly drops as the wind expands outward.
Radiative losses are negligible for
the high specific energy winds simulated here, so this
scenario is relevant to our simulations. However,
while this conversion of thermal to kinetic energy
is significant at small radii for the \highark{}
simulation, we can see that for the \highspec{}
case (which lacks the novel recoupling and
refinement scheme) the balance of energy components
remains close to the injected ratio. By $0.2~r_{200}$
the \highspec{} has converted some of its thermal
content into kinetic energy, but not to the extent
of the \highark{} simulation at the same radius.

We note that for both high specific energy winds the
peak value of $\dot{E}_\mathrm{kin}/\dot{E}_\mathrm{th}$
increases marginally as a function of radius until 
$\sim0.5~r_{200}$. 
However, while the value is relatively constant in time
at $0.1~r_{200}$, further out there is a gradual decline.
This is related to the narrowing of the wind region as its
power decreases with falling SFR, as described above.
The geometry of the flow is important for the amplitude
of the thermal to kinetic energy conversion. As the
wind narrows, it transitions from a geometry
where flow lines spread out as a function of radius
to an arrangement where the flow lines
are more or less parallel as the wind moves primarily
vertically upwards away from the disc plane.
As shown in \cite{Martizzi2016}, a cartesian
geometry lacks the $1/r^2$ spherical divergence
term in the \citetalias{Chevalier1985} solution
that permits the formation of supersonic winds.
Our case is not so severe, since streamlines
can open up at small radii where the majority
of the acceleration occurs, so our winds go
through a sonic point (as we shall show below).
However, the last stages of the thermal to kinetic
energy conversion are blunted as the outflow
becomes restricted by the CGM at larger radii.
Finally, at late times when the geometry
is at its most ``cartesian'', the value
of $\dot{E}_\mathrm{kin}/\dot{E}_\mathrm{th}$
at large radii drops below its central peak.
This is due to the wind being decelerated
by gravity, which
the aforementioned acceleration had
previously balanced.

\subsubsection{Radial profiles of relevant quantities} \label{sec:profiles}

In \cref{fig:profiles_subsonic_5mdisk} we show various
quantities as a function of galactocentric radius at various
times,
measured within a 
bicone with opening angle $10\degree$ orientated along
the rotational axis of the disc/CGM (i.e. the initial
wind launch direction). We show results from the
\subfive{} no wind, \lowspec{}, \highspec{} and \highark{}
simulations.
Quantities are measured in a similar
manner to the method we use for calculating outflow rates
in the preceding section. At each radius, we discretise
the spherical cap enclosed by the cone opening angle 
into equal area pixels using \textsc{HealPix}. The properties
of the nearest gas cell (mesh-generating point) are mapped
onto the pixels. We then report the mean value of those
properties across all pixels at a given radius.\footnote{These
are therefore solid-angle weighted averages. We find that
density weighting the cell contributions yields almost
identical results.}
From top to bottom, the rows of 
\cref{fig:profiles_subsonic_5mdisk} show
density, pressure, sound speed and radial velocity, and cell radius,
respectively.

The \lowspec{} wind shows a noticeable density enhancement
relative to the no wind case in all but the very central
region. The high mass loading of the wind leads to an
excess of material, as seen in \cref{fig:slice_subsonic_5mdisk},
particularly within $0.1~r_{200}$ ($\sim$10~kpc). In fact, at
2~Gyr this results in all gas within $\sim4.5$~kpc in the
bicone being above the wind particle recoupling density threshold, shown
as a dotted gray line. By 0.5~Gyr the wind is only travelling outwards
at $\sim100~\kms$, coasting from the initial burst. At later times,
it can be seen that the bulk velocity changes sign (indicating inflow)
at various radii. The wind is of a lower temperature than the surrounding
CGM. The initial thermal component, giving wind particles a subgrid
sound speed of 94.1~$\kms$ ($T=3.8\times10^5\,\mathrm{K}$), is
quickly radiated away. This means that, despite the low velocities,
the inflows and outflows are both supersonic on the whole.
Finally, in the bottom row, we see that cells in
the wind have sizes that are small compared to their galactocentric radius,
the relatively high density translating to a correspondingly fine spatial
resolution. Note that at the very centre (as we enter the disc region),
the cell size is comparable to the galactocentric radius. In combination
with the small spatial extent of the cone at this distance, this means
that the profiles are now tracing individual cells.

Turning to the high specific energy winds, we see some features in
common between the \highspec{} and \highark{} simulations. As shown
in \cref{fig:slice_subsonic_5mdisk}, the winds have a much lower
density, higher pressure, higher sound speed and higher velocity than in the \lowspec{}
simulation. However, as expected from 
\cref{fig:load_relative_subsonic_5mdisk}, we can see that the
\highark{} winds reach significantly higher velocities than the
\highspec{} case at all radii and times. The \highark{} wind is initially
mildly subsonic at small radii, but rapidly accelerates, passing through
a sonic point at $R\sim1\,\mathrm{kpc}$. 
Beyond the sonic point, it largely follows the \citetalias{Chevalier1985}
asymptotic radial evolution ($\rho \propto r^{-2}$,
$P \propto r^{-10/3}$, $v = \mathrm {const.}$) to a few 10s of kpc.
We stress that the \citetalias{Chevalier1985} solution is 
only approximately applicable to our setup as we include a central galaxy,
a CGM, radiative cooling and gravity. However, it is instructive
that in the regime where these differences have the least importance,
where the wind is essentially expanding adiabatically, our
scheme produces a wind with comparable evolution.

By contrast, the \highspec{}
simulation remains subsonic until much further out ($\sim5-7\,\mathrm{kpc}$)
 and undergoes
a more gentle acceleration. In the very central region the sound speed
is lower, since wind particles inject their energy into
a larger mass in the absence of the new recoupling and refinement scheme.
Note also
the large spikes of pressure at 1 and 2~Gyr, corresponding to recent
wind particle recouplings due to the poor temporal energy resolution
(see \cref{sec:temporal_res}).
However, from a few kpc outwards the sound speed
is higher than the \highark{} wind as thermal energy is not converted into kinetic energy as
rapidly (consistent with \cref{fig:load_relative_subsonic_5mdisk}). 

The reason for the contrasting behaviour lies in the difference between
the typical cell size in the two winds. Due to the low densities in the
wind region, the cells in the \highspec{} simulation are very large,
often comparable to their galactocentric radii. In other words, the
central regions of the wind (where \citetalias{Chevalier1985}
predict there should be the sudden conversion of thermal to kinetic
energy) is only spanned by a handful of cells. This can also be seen visually
in \cref{fig:slice_subsonic_5mdisk}. This means that the
relevant scales\footnote{The most intuitive comparison is to the
galactocentric radius, $r$, which is what we show in \cref{fig:profiles_subsonic_5mdisk}.
However, the gradient length scales of density, pressure, velocity etc. are also
relevant. In the \citetalias{Chevalier1985} solution, these are generally
within a factor of a few of $r$ for parameters close to our setup. 
Thus, comparing
cell size to $r$ is a convenient proxy when assessing how well resolved the flow is.}
are poorly resolved exactly where the most important part of the
wind's radial evolution is located. This leads to a much gentler acceleration
of the wind material.
A second related impact of the poor resolution is that the minimum size of
the injection region is necessarily comparable to the typical cell
size. Ideally, since the energy injection in our model is intended to 
represent the emergence of wind material out of the ISM, this injection
region should be as small as possible. The coarse spatial resolution
in the \highspec{} simulation means that the effective injection region
is much larger. This leads to the sonic point being displaced
to larger radii. The combination of these two resolution effects
means that failing to properly resolve the low density, high specific
energy gas at the base of the wind leads to a lower acceleration
of material and a larger retention of thermal energy.

Interestingly,
as shown in \cref{fig:mflows_subsonic_5mdisk,fig:load_relative_subsonic_5mdisk},
despite these radical differences, the net mass and energy fluxes
between the two runs are very similar. This is because both
winds have comparable mass and energy injection rates (due to 
having the same loadings and similar SFRs). Energetic losses
are negligible, due to the high velocities and long cooling times
involved, so the outflowing mass and energy fluxes are
conserved, leading to convergence in those bulk properties
between the two simulations. \highspec{} may drive slower winds,
but it is denser such that the mass outflow rate is similar to
the \highark{} simulation. Likewise, the net energy fluxes
are comparable despite the different partitioning into kinetic
and thermal components.

\subsubsection{Temporal evolution of the wind opening angle}
\begin{figure} 
\centering
\includegraphics{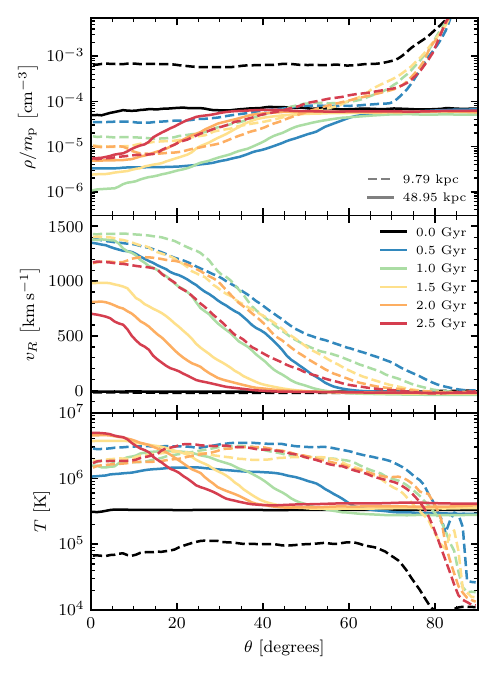}
\caption{For the fiducial \highark{} simulation, we plot density (top row),
radial velocity (middle row) and temperature (bottom row) as a function
of angle from the rotation axis of the disc/CGM. Dashed (solid) lines show the
measurements at 0.1~$r_{200}$ (0.5~$r_{200}$). Different times throughout
the simulation are shown with different colours. The region occupied by the
wind (with low densities and high velocities and temperatures) narrows
as the simulation progresses.
}
\label{fig:angle_profiles_onesim_subsonic_5mdisk_highspec_ark} 
\end{figure}

In \cref{fig:angle_profiles_onesim_subsonic_5mdisk_highspec_ark}
we show gas density, radial velocity and temperature as a function
of angle, $\theta$, away from the rotation axis of the disc/CGM (which is also
the axis of the wind) at 0.1 and 0.5~$r_{200}$ for the \highark{} simulation
at various times.
Quantities are measured in a comparable manner to the previous section.
The regions above and below the disc plane are mirrored onto a single
hemisphere. As was shown in \cref{fig:slice_subsonic_5mdisk},
the wind region is well defined, with the lowest densities and
highest velocities and temperatures occurring on-axis, before returning
to an undisturbed CGM (similar to those in the initial conditions i.e.
$t=0~\mathrm{Gyr}$ with increasing $\theta$).
As remarked upon earlier and
shown in \cref{fig:slice_dens_subsonic_5mdisk_highspec_ark},
the wind becomes narrower with time. This is particularly noticeable
at $0.5\,r_{200}$, where the width of the wind region decreases
considerably over the 2.5~Gyr of the simulation. As it does so,
the velocity of the wind drops at all angles. This is less true
of the measurements taken at $0.1\,r_{200}$, which show much less
evolution in both the opening angle of the wind and the velocity.
At this small distance, the CGM plays less of a role in confining
the wind.

\subsection{Vertical vs. Isotropic launching} \label{sec:direction}
\begin{figure} 
\centering
\includegraphics{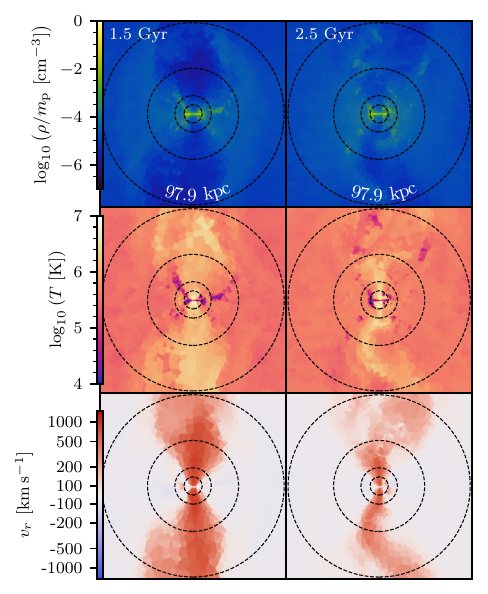}
\caption{Slices showing density (top row), temperature (middle row)
and radial velocity (bottom row) for the fiducial ICs with
the isotropic \highark{} wind model. The left (right) column
shows the simulation at 1.5~Gyr (2.5~Gyr).
The general morphology is similar to the fiducial \highark{}
simulation, with a biconical outflow developing despite
the isotropic injection. The wind is more disturbed, however,
with clumps of ISM material being expelled from the edges of the
disc and entrained along the edge of the wind.}
\label{fig:slice_subsonic_5mdisk_highspec_ark_iso} 
\end{figure}

\begin{figure*} 
\centering
\includegraphics{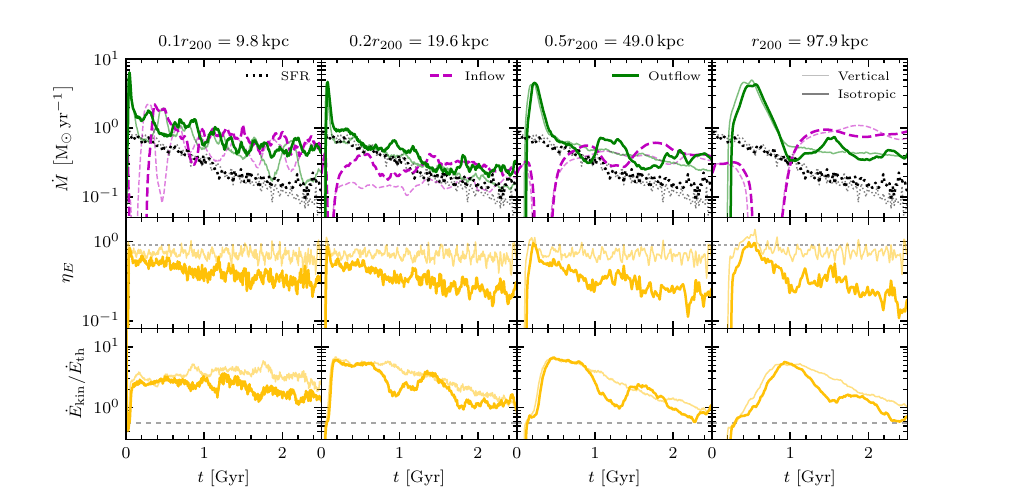}
\caption{Mass fluxes (top row), energy loading (middle row) and
kinetic to thermal energy flux ratio for the fiducial ICs
\highark{} simulation with isotropic wind particle launching. The 
equivalent simulation with vertical launching (shown
throughout \cref{sec:fid_sims}) is shown with a thin
line for reference. Columns show the measurements at different 
galactocentric radii. Isotropic and vertical launching
produce very similar results. The emergent energy loadings
in the isotropic case drop as a function of time, compared to the
vertical launching simulation which remain flat. This is because
the energy is diluted over a wider solid angle, which causes greater
losses particularly as the absolute energy input reduces with
declining SFR. The wind also has more cold material being mixed into
it than the vertical case.}
\label{fig:mflows_and_load_subsonic_5mdisk_iso} 
\end{figure*}

We mentioned in \cref{sec:particle_creation} that our implementation
allows wind particles to be launched either vertically out of the disc plane
or isotropically. In the isotropic case, wind particles may be launched
in the ``wrong'' direction; for example, there is an equal probability that
an ISM gas cell sitting above the disc plane will launch a particle downwards
into the disc as upwards. In this case, however, the particle should travel
through the entire disc without triggering the density-based recoupling also occurs for gas cells below the disc, the net flux of particles
balances. However, particles can be launched along trajectories 
near-parallel to the disc plane,
causing them to travel significant distances through dense gas.
Isotropic launching in principle has two advantages over vertical
launching in the case of cosmological simulations.
Firstly, it requires no knowledge of a preferred
direction, removing the need to determine this on the fly.
Secondly, it naturally copes with situations where there is
no clear disc where an isotropic outflow is likely the more
physical outcome. 
It should be noted, however, that if properly constructed, 
a scheme which
uses a preferred launch direction could smoothly transition
between isotropic and vertical launching depending on
the presence or absence of a disc-like configuration
(e.g. launching along the potential or density gradient)
but would be more complex. We leave such an exploration to
a future work.

The vertical scheme is our fiducial choice in this work, but
we now show results of rerunning the \subfive{} \highark{} simulation
with isotropic launching.
\cref{fig:slice_subsonic_5mdisk_highspec_ark_iso} shows slices
of density, temperature and radial velocity at 1.5 and 2.5~Gyr for
this simulation. Despite launching wind particles isotropically,
it can be seen that the wind still emerges as a biconical outflow
flowing vertically away from the disc plane. Qualitatively, at 1.5~Gyr,
the morphology is comparable to the vertical launching case shown in
\cref{fig:slice_subsonic_5mdisk}.
However, unlike the vertically launched case, there are cold, dense
clumps of material present near the edges of the outflow at
1.5~Gyr. In the 2.5~Gyr image, clumps can be seen mingled in
with the body of the wind, causing diversions to the flow and
a more disturbed morphology. These clumps of material are
stripped from the disc primarily from the outer regions, by wind particles
launched with oblique angles relative to the disc plane. They recouple
within the disc when they encounter low density material. This is
either because they have travelled to the edges of the disc where the
density naturally declines or they have arrived in a low density
cavity. They then inject their energy into this gas, which disrupts
the disc and expels ISM material. 
This process efficiently leads to the removal of the outer portions
of the disc.
ISM material expelled in this manner
typically remains confined to the edges of the wind, trapped between
the main body of the outflow and the surrounding CGM (see the slices
at 1.5~Gyr). Occasionally, clumps end up in the main body of the
wind, usually because they were lifted from a more central region
of the disc, where they disrupt the outflow (see the slices
at 2.5~Gyr) before being shredded.

\begin{figure*} 
\centering
\includegraphics{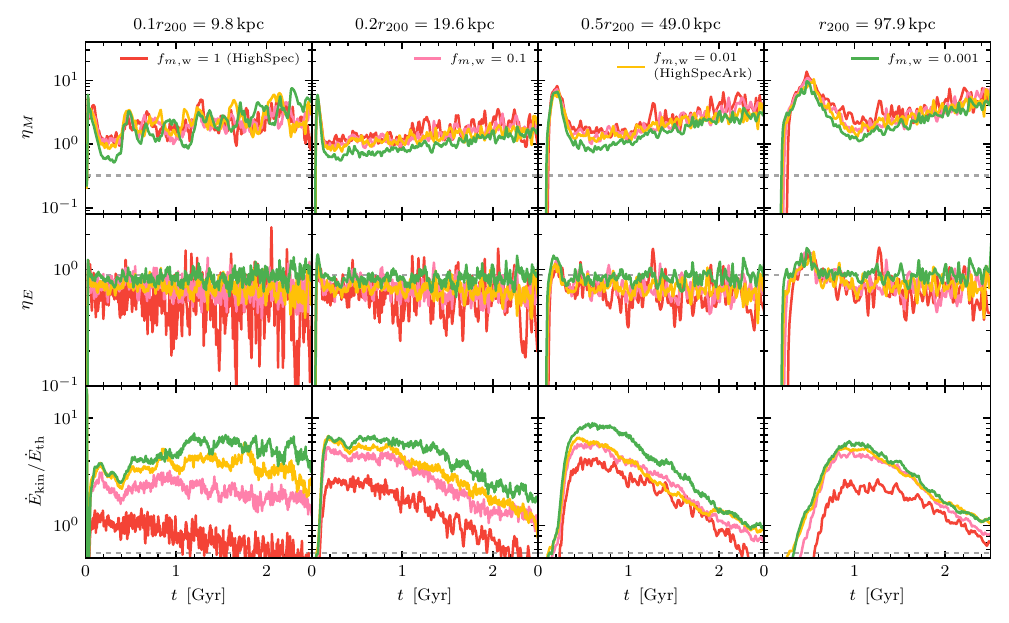}
\caption{Mass loadings (top row), energy loadings (middle row)
and ratio of kinetic to thermal energy fluxes (bottom row)
for the fiducial ICs with high specific energy winds 
with different values of $f_{m,\mathrm{w}}$. Columns from left to right show 
measurements at 0.1, 0.2, 0.5 and 
1~$r_{200}$, respectively.
The $f_{m,\mathrm{w}}=1$ case is the \highspec{} simulation presented
in the earlier parts of this work; it does not use the novel 
displacement recoupling and
refinement scheme. The remaining simulations shown do use the new
scheme. The $f_{m,\mathrm{w}}=0.01$ is the fiducial \highark{} simulation
previously presented. Average mass and energy loadings are
very similar among all the simulations. However, at lower refinement
levels the energy loading is much burstier. More strikingly,
the ratio of kinetic to thermal energy increases with better resolution,
although the difference between the highest resolution runs is small.}
\label{fig:load_relative_subsonic_5mdisk_resolution} 
\end{figure*}

In \cref{fig:mflows_and_load_subsonic_5mdisk_iso} we
show mass fluxes (SFR, inflows and outflows), energy loadings
and kinetic to thermal energy flux ratio for the isotropic
simulation, similar to \cref{fig:mflows_subsonic_5mdisk,fig:load_relative_subsonic_5mdisk}. 
We also show the
fiducial vertical launched \highark{} simulation, for reference,
with thin lines.
In general, the two simulations give very similar results.
In particular, the mass fluxes are near identical at all
radii (inflows are marginally enhanced at 0.2~$r_{200}$
in the isotropic case). The energy loadings are marginally
lower in the isotropic case and decline over time.
This is a consequence both of the dilution of energy over
a wider volume (a certain fraction of the energy ends up
being injected towards the edge of the disc, rather than
all being coupled directly above it) and of the interactions
with the cold, dense clumps. The partition of energy into
kinetic and thermal components is similar at all radii, with
marginal differences, matching expectations given the
aforementioned difference in injection region and losses
to swept up ISM clumps.

As we also find, \cite{Pillepich2018} show that they drive non-spherically symmetric
outflows even though they launch their wind particles isotropically,
because the flows naturally move along the path of least resistance.
However, their gas morphologies do not appear to be substantially different
from the case where they launch their winds out of the disc plane.
In our case, the disruption of parts of the disc when we launch
isotropically is likely because our higher specific energy
wind particles can have a much larger impact when they find a low
density region within the ISM. This suggests that a fully isotropic launching
scheme may not be appropriate when using high specific energy
winds. When we launch our \lowspec{} wind isotropically
(not shown) we see little additional disruption of the ISM beyond that
which occurs from mass loading the wind. The only major difference is the
lack of the initial highly collimated vertical plume, with only low
altitude fountain flows present.

\subsection{Refinement level and computational expense} \label{sec:resolution}
The \highspec{} simulation presented in previous sections launches
wind particles that have the same mass resolution as the target gas
mass resolution of the simulation (in our notation, this
is $f_{m,\mathrm{w}}=1$, see \cref{eq:f_mw}), which
is $8\times10^4\,\Msun$.
Because the wind particle masses match those of ambient gas and the specific energy is high,
where they recouple the cell temperature is significantly raised;
there would be no benefit from applying our new displacement recoupling scheme.
The \highark{}
simulation launches wind particles that are 100 times less massive
than the target gas mass resolution ($f_{m,\mathrm{w}}=0.01$),
$800\,\Msun$. It uses the full \textsc{Arkenstone} recoupling and 
refinement scheme. We have repeated the \highark{} with coarser and
finer mass resolution in the wind particles/hot wind cells in order to
examine the impact of this choice. 
These variation runs use $f_{m,\mathrm{w}}=0.1$ and $f_{m,\mathrm{w}}=0.001$
i.e. a wind mass resolution of $8000\,\Msun$ and $80\,\Msun$, respectively.

\cref{fig:load_relative_subsonic_5mdisk_resolution} shows the mass loadings,
energy loadings and kinetic to thermal energy flux ratios for these runs
(analogously to \cref{fig:load_relative_subsonic_5mdisk}). The mass loadings
and energy loadings are very similar between the four runs, despite the three
orders of magnitude mass resolution difference spanned. This is unsurprising in
the context of the results presented earlier; as we remarked in
\cref{sec:profiles}, energy losses in the wind material are negligible,
so mass and energy fluxes are conserved, producing the same bulk properties
in all runs. The burstiness of the energy fluxes, particularly at small scales,
does depend on wind particle mass. We see that while the $f_{m,\mathrm{w}}=0.1$
is slightly more bursty than our fiducial case ($f_{m,\mathrm{w}}=0.01$),
the difference is marginal compared to the large amplitude fluctuations that
occur when no refinement is used.

More significantly, the partition of energy into thermal and kinetic components (i.e. the
degree to which the gas is accelerated) is resolution dependent, as we described
in previous sections. Here differences are apparent, with finer resolution yielding a
higher value of $\dot{E}_\mathrm{kin}/\dot{E}_\mathrm{th}$.
Nonetheless, increasing the degree of refinement by an
order of magnitude (green) from our fiducial case yields at most differences of a factor of 1.3 - 1.4
at $0.1\,r_{200}$ compared to the almost order of magnitude discrepancies when comparing
\highark{} to \highspec{}. If we instead reduce the level of refinement, comparing the
$f_{m,\mathrm{w}}=0.1$ to our fiducial \highark{}, we see the degree of acceleration is
indeed smaller, as expected, but again the differences are slight even at small radii.
This suggests that, while the simulations using our new recoupling and refinement scheme
are evidently not completely converged in this
respect as a function of $f_{m,\mathrm{w}}$, we might expect that they are not far off,
particularly when comparing to the simulation that does not use any refinement.

While it has been convenient to parameterise the resolution \textit{relative} to the
base resolution of the simulation (i.e.
with the $f_{m,\mathrm{w}}$ parameter), it is important to note that the resolution
requirements are \textit{absolute}.
As described above, the key requirement is that
the simulation has sufficient spatial resolution near the base of the wind as
it is accelerated through the sonic point. This, therefore, ties the resolution
requirement to the properties of the wind being simulated. Based on the results
in \cref{fig:load_relative_subsonic_5mdisk_resolution}, we suggest that
$800\,\Msun$ in the wind particles/hot wind cells yields near optimal results
for this wind,
though $8000\,\Msun$ is likely sufficient. More aggressive refinement yields
diminishing returns. However, if we coarsened the base resolution of the simulation
we would have to reduce $f_{m,\mathrm{w}}$ accordingly to achieve the same absolute
resolution in the wind. We demonstrate this in \cref{ap:base_res}, showing
that the \textsc{Arkenstone} scheme functions well in simulations with
a coarser base resolution.
We also stress that these resolution requirements are
specific to the configuration of this wind (e.g. mass and energy loadings, SFR
of the galaxy etc.). Care must be taken, therefore, to adjust the value of
$f_{m,\mathrm{w}}$ according to circumstances. If the cell size is comparable to the
galactocentric radius as the wind goes through its sonic point the wind
acceleration is likely under-resolved.

At this point, it is worth reporting the additional computational expense
associated with the displacement recoupling and refinement scheme, relative
to the \highspec{} simulation which does not use them. 
First of all, it is worth remarking that all of the simulations presented
in this work required negligible computational resources by the standards
of most contemporary simulations of individual galaxies, because we intentionally
only used a resolution appropriate for a cosmological volume. The
\highspec{} simulation required only 126~CPUhr to reach 2.5~Gyr of evolution.
Our fiducial \highark{}
simulation, which refines the wind by a factor of 100 in mass, increases runtime
by a factor of 1.3.
The variation runs in this section which refine
by a factor of 10 and 1000 are 1.1 and 3.3 times more expensive than the \highspec{}
simulation, respectively.
The extra computational expense is not directly associated with the operation of the scheme
itself (e.g. neighbour searches, refinement operations etc.) which makes up approximately
0.2 per cent of the total cost of the simulation. The scheme results in an increased
population of cells on shorter timesteps because cells near the base of the wind are in
general hotter and have smaller diameters. This partially contributes to the additional
cost.

The other additional expense is associated with derefinement operations,
which occur more frequently when the scheme is active. Due to the unstructured
mesh used in Arepo, derefining a cell incurs computational costs associated
with reconstructing the local mesh and remapping conserved quantities onto neighbouring
cells
\citep[for details, see][section 6]{Springel2010}. 
Derefinement accounted for 1.5\% of the runtime of \highspec{}, rising to
8.4\% for \highark{} and 17.4\% for the variation of \highark{} which refines
by an additional factor of 10.
We chose a relatively
arbitrary derefinement criterion for the wind (i.e. a return to the target mass
of the simulation as a linear function of galactocentric radius between
$0.1r_{200} - 0.5r_{200}$). Alternative criteria could in principle
be found, balancing more efficient derefinement operations against
the additional cost of maintaining a higher resolution over a greater
volume. However, we felt that it was not worth pursuing further optimisation
until the scheme is deployed in its intended cosmological setting.
It should be noted that, in general, the relative expense
of the scheme depends not only on the refinement level and input mass and energy loadings (these
combine to determine the number of wind resolution elements and their timesteps) but also
the fraction of expense associated with other components
(non-wind gas, stars, dark matter etc.). Furthermore, the relative expense
of different parts of the simulation (e.g. gravity, hydrodynamics, mesh construction etc.)
will likely be very different in a cosmological volume, so we must avoid
overinterpreting these findings.

\subsection{A more massive CGM: Supersonic ICs}\label{sec:supersonicICs}

\begin{figure} 
\centering
\includegraphics{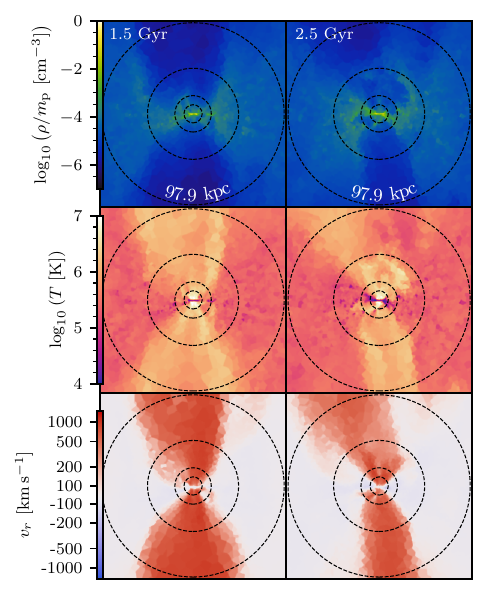}
\caption{Slices showing density (top row), temperature (middle row)
and radial velocity (bottom row) for the Supersonic ICs with
the \highark{} wind model. The left (right) column
shows the simulation at 1.5~Gyr (2.5~Gyr). The general
morphology is similar to the fiducial case, but the inflow
from the CGM is more clumpy. This occasionally disrupts the
outflow.}
\label{fig:slice_supersonic_highspec_ark} 
\end{figure}

\begin{figure*}
\centering
\includegraphics{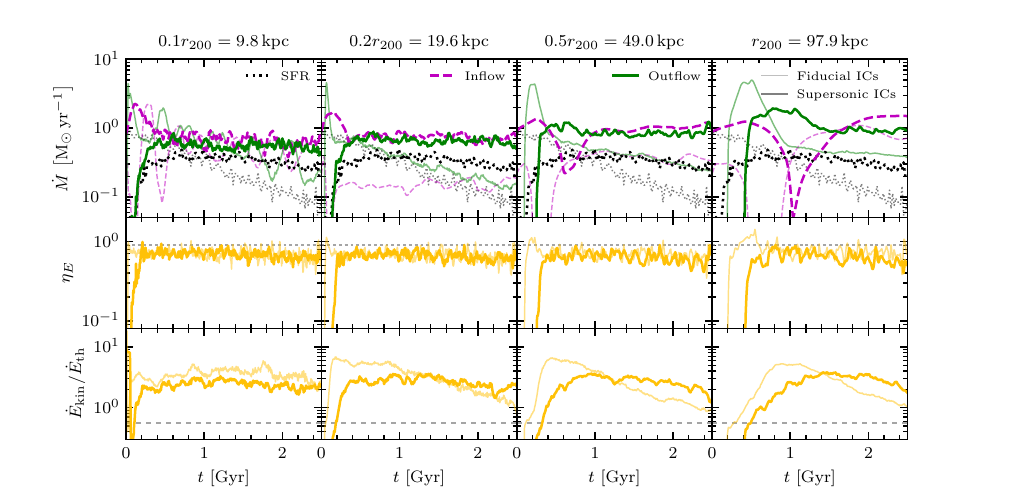}
\caption{Mass fluxes (top row), energy loading (middle row) and
kinetic to thermal energy flux ratio for the Supersonic ICs
\highark{} simulation. The fiducial ICs \highark{} is shown
with a thin
line for reference. Columns show the measurements at different 
galactocentric radii. The Supersonic ICs have a lower
disc mass (by a factor of 5) which leads to a smaller
initial SFR. However, the inflows and outflows are comparable at
all radii, leading to a constant SFR as the disc can be replenished,
in contrast with the fiducial ICs simulation.}
\label{fig:mflows_and_load_supersonic} 
\end{figure*}

Here we examine the impact of starting the simulation
with a lower galaxy mass, but a stronger cooling flow,
corresponding to a more massive CGM.
These are the ``Supersonic ICs'' detailed in \cref{tab:IC}.
The masses of the stellar disc and bulge 
and the gas disc in these initial conditions are a factor of five lower
than the fiducial simulations shown above, but have the same scale lengths.
We use a cooling flow solution that yields a
supersonic flow in the inner regions, with $r_\mathrm{sonic}=5\,\mathrm{kpc}$
compared to $R_\mathrm{circ}=2.5\,\mathrm{kpc}$ (the latter matching the
scale length of the gas and stellar discs). The solution gives a
CGM that is a factor of 1.46 more massive inside $r_{200}$ and
predicts a mass inflow rate around four times higher than our fiducial initial
conditions.
We rerun the \highark{} model with these initial conditions.

\cref{fig:slice_supersonic_highspec_ark} shows slices of the
simulation at 1.5 and 2.5~Gyr. Compared to the fiducial simulations,
the denser CGM is apparent outside of the outflowing regions,
particularly as it flows into the galaxy, aligned with the disc
plane. At some points in the simulation, such as at 1.5~Gyr, the
wind morphology is qualitatively similar to our fiducial case,
with a large symmetric biconical outflow with a large opening
angle. However, at other times the clumpy inflowing medium
close to the galaxy marginally disrupts the outflow, leading
to a more irregular structure (for example, as shown at
2.5~Gyr).

\cref{fig:mflows_and_load_supersonic} shows mass fluxes,
outflow energy loadings and the ratio of kinetic to thermal
energy fluxes in the wind. The fiducial \highark{} simulation
is plotted with thin lines for reference. The galaxy does not
experience the burst of SFR that occurs in the fiducial
simulation, due to its lower gas surface densities. It therefore
takes $\sim$200-400~Myr before outflows are established.
Once this occurs, the inflow and outflow rates effectively
balance at all radii. However, the winds do not prevent inflowing
material from reaching the disc to the extent that occurs in the fiducial
simulation, with an efficient inflow present in the disc plane. The
result is that the galaxy maintains a steady SFR, despite
having an initially smaller gas reservoir. Accordingly,
the wind achieves a steady state, with constant
mass and energy fluxes. This means that the narrowing
of the wind region that occurs in the fiducial simulations,
a consequence of declining wind power, does not happen
in this simulation. The interactions with dense clumps
can cause temporary fluctuations in the morphology
(such as that shown in the 2.5~Gyr images) but
the opening angle of the wind is not reduced in the
long term. These differences aside, it can be seen
in \cref{fig:mflows_and_load_supersonic} that
the behaviour with respect to emergent energy loadings
and energy partition is qualitatively the same as the
fiducial case. The wind preserves its initial energy
loading as it flows out of the halo and experiences
the same rebalancing of thermal to kinetic energy
at small galactocentric radii.

\subsection{Medium specific energy loadings} \label{sec:twindfix}
\begin{figure} 
\centering
\includegraphics{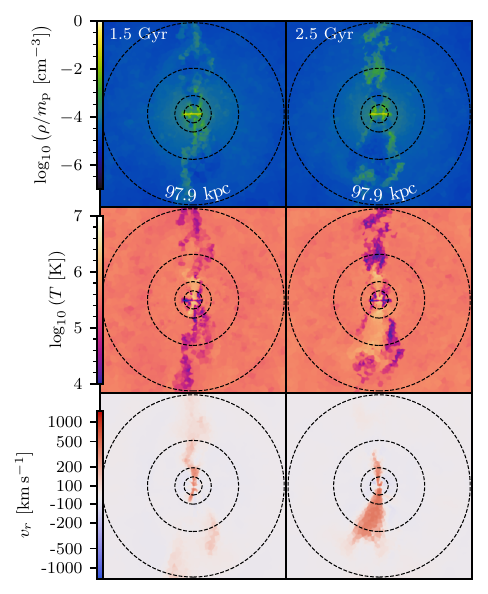}
\caption{Slices showing density (top row), temperature (middle row)
and radial velocity (bottom row) for the fiducial ICs with
the \medark{} wind model. The left (right) column
shows the simulation at 1.5~Gyr (2.5~Gyr).
Compared to the fiducial \highark{} simulations, outflows are
weaker and emerge predominantly from the central region of the disc.
The wind does not fill such a large volume and has a more
disturbed morphology as it interacts with the CGM and turns around in places.}
\label{fig:slice_subsonic_5mdisk_twindfix} 
\end{figure}

\begin{figure*} 
\centering
\includegraphics{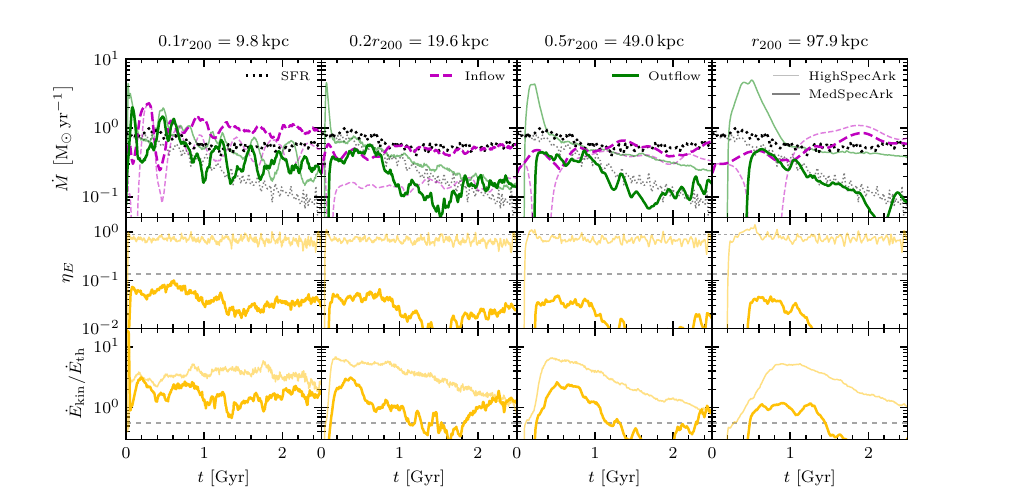}
\caption{Mass fluxes (top row), energy loading (middle row) and
kinetic to thermal energy flux ratio for the fiducial ICs
\medark{} simulation. The \highark{} simulation is shown
with a thin
line for reference. Columns show the measurements at different 
galactocentric radii. Outflows rates are not as high
as the \highark{} simulation, particularly at larger radii,
and inflows are not so suppressed, leading to a high SFR.
The input energy loading is not preserved. However, the general
feature of rebalancing thermal to kinetic energy is present.}
\label{fig:mflows_and_load_twindfix} 
\end{figure*}
Finally, we perform a simulation 
with our \subfive{} initial conditions with a wind with
the \medark{} wind model (see \cref{tab:wind}).
This uses the full displacement recoupling and
refinement scheme.
It uses input loading factors of
$\eta_M=0.24$ and $\eta_{E,\mathrm{tot}}=0.135$ (with thermal and kinetic
energy loadings balanced to give $\mathcal{M}_\mathrm{launch}=1$).
These loadings are derived from the \cite{Kim2020a} relations
for the total hot component for 
$\Sigma_\mathrm{SFR}=6\times10^{-2}\,\Msun\,\mathrm{kpc^{-2}}\,\mathrm{yr^{-1}}$
(which is approximately the initial value of $\Sigma_\mathrm{SFR}$ inside
one scale radius in our initial conditions). This mass loading is similar
to that used in our fiducial high specific energy simulations, but with
an energy loading almost a factor of seven lower. We stress
that this simulation should not be interpreted as an application of
the \cite{Kim2020a} \textsc{Twind} wind launching model (this has been
implemented in \textsc{Arkenstone} and will be explored in
a subsequent work). Firstly, a key aspect of the \textsc{Twind} model
is that it produces distributions of temperatures and velocities within
a given wind component (rather than the single total loadings we use here).
Secondly, we use fixed loadings that do not adjust adaptively in space
and time to reflect local ISM conditions. Finally, we only include the hot
wind component here, ignoring the much more mass loaded cold component.
Taken together, these aspects mean that the full \textsc{Twind} model
leads to a significantly more complex mode
of wind driving. However, it suits our purposes in this work to use the
loadings described above to examine the evolution of a hot wind component
with a lower specific energy than our fiducial case.

In \cref{fig:slice_subsonic_5mdisk_twindfix} we show slices of
density, temperature and radial velocity for this simulation at
1.5 and 2.5~Gyr. In \cref{fig:mflows_and_load_twindfix}
we show mass fluxes (inflow, outflow and SFR),
emergent energy loadings and the ratio of kinetic to
thermal outflowing energy fluxes.
Due to its lower power, 
the wind is less able to push outwards against the inflowing CGM
than the higher specific energy cases shown above. This results
in a highly confined outflow. Additionally, outflows
are predominantly launched from the centre of the
disc. The wind mass and energy flux per unit area launched from the disc is proportional to the
SFR surface density, $\Sigma_\mathrm{SFR}$. This means
that the wind is not as strong towards the edges of the
disc where $\Sigma_\mathrm{SFR}$ declines (due
to dropping gas surface density).
This hinders the ability to drive effective winds into the CGM from the outer regions
of the disc. This phenomenon was not encountered in the
\highark{} simulation because of the significantly
higher input energy loading. This further compounds
the confinement issue described above, leading to a
much smaller wind opening angle.
The wind is therefore unable to suppress inflow rates
to the same extent is the \highark{} simulation,
as demonstrated in \cref{fig:mflows_and_load_twindfix}.
Accordingly, the SFR remains relatively steady due
to the replenishment of the ISM.

Due to the reduced wind opening angle, inflowing material
can more readily approach the disc at steeper angles relative
to the disc plane. This material often
passes into the wind region itself, causing a
disruption to the outflow (in a similar manner to
that seen in \cref{sec:supersonicICs}) which
contributes to the
complex morphology visible in \cref{fig:slice_subsonic_5mdisk_twindfix}.
In addition to this first-inflow CGM material, the wind is also
hindered by the return of gas launched from the galaxy that
has stalled and turned around. 
This is not nearly as severe
as the situation experienced by the fiducial \lowspec{}
simulation, with its almost 27 times higher mass loading, but
shares similar characteristics. 
As can be seen in \cref{fig:mflows_and_load_twindfix}, a significant
fraction of the mass launched outwards through 0.1~$r_{200}$
does not make it out of the halo. This material builds up
in the outer halo and gathers into regions of over-dense,
cooling gas. The gas then falls back towards the galaxy
and interacts with outflowing material.

Examining the emergent energy loadings in \cref{fig:mflows_and_load_twindfix},
we see that the input loadings are not preserved, in contrast
to the \highark{} simulation. The energy loading drops with
increasing distance. At large radii, the decrease in energy loading
largely reflects the reduction in wind material actually travelling
that far. This is a consequence of energy losses at smaller radii.
These come from deceleration by gravity (as a fraction
of the initial kinetic energy, the lower velocity wind
of \medark{} suffers from this more than the higher
velocity wind of \highark{}), from the interactions with the CGM
and from cooling (the wind starts out at a lower temperature
than \highark{} and has higher density structures, for reasons
explained above). Despite the increased energy losses,
the \medark{} simulation still shows the same qualitative
evolution of the kinetic to thermal energy flux ratio as
the \highark{} case, at least at small radii. It can be
seen that the re-balancing of thermal to kinetic energy
is present. This again demonstrates the importance of
resolving the wind properly as it passes through
its sonic point, even for these lower energy loadings.

\section{Discussion} \label{sec:discussion}
\subsection{Resolving high specific energy winds}
As we have shown above, there are several resolution-related
challenges inherent to modelling
high specific energy (i.e. hot, fast) winds in simulations of galaxy
formation. Firstly, it can be difficult to inject energy smoothly
since energy resolution is often tied to mass resolution.
We demonstrated that this is the case when using wind particles
at coarse resolution. However, this link also exists in
schemes that inject energy directly into the ISM. For example,
in the stochastic heating scheme of \cite{DallaVecchia2012},
the requirement that injection events produce a sufficient
temperature increase results in increasingly rare injections
of large quantities of energy, for the same SFR, as resolution
is coarsened.
On small scales, the efficient driving of
galactic winds with stellar feedback is likely associated with
clustered SNe, working together to create superbubbles
\citep[see e.g.][]{Yadav2017,Kim2017b,Fielding2018,Gentry2019,Smith2021a}.
This might suggest that a
highly bursty injection of energy is a desirable feature
of a galactic wind model. However, it is important to draw a
distinction between this physical clustering, and artificial
burstiness derived from numerical stochasticity, which are
unrelated. It is crucial, therefore, to discretise the
injection of energy as finely as possible.
Even better energy resolution is required if wind particle
velocities and temperatures will be sampled from
a distribution \citep{Kim2020a}, a capability of
\textsc{Arkenstone} which will be explored in future work.

Simply increasing the mass (and therefore energy) resolution
of wind particles would not, by itself, solve the issue because this
would lead
to overcooling as the energy would be lost to cooling
when it is injected into coarse resolution
gas in a piecemeal fashion.
Our displacement recoupling scheme
tackles this obstacle by increasing mass resolution
at the point of injection. This solves the overcooling problem
with the opposite approach to stochastic heating feedback, which
achieves high specific energies by injecting large quanta of
energy into large amounts of mass; we gain the same high specific
energies by injecting smaller quanta of energy into less mass.
At coarse resolution, it becomes harder to avoid coupling
feedback energy into an unphysically large quantity of mass.
Our method, the redistribution of mass in the ISM/CGM transition
region, aims to compensate for the inability to resolve the
venting of hot wind material out of the ISM via bubbles and
chimneys.

The higher resolution provided by the displacement recoupling
scheme helps assist the energy injection, as described above,
but maintaining it with the modified refinement scheme
is also crucial for capturing the properties of the wind
as it flows away from its source. As we showed with our simulations,
a lack of spatial resolution (resulting from fixed cell mass
in low density material) leads to incorrect partitioning of energy
between kinetic and thermal components. 
Without sufficient resolution,
the rapid acceleration
of the flow through a sonic point \citepalias[see e.g.][]{Chevalier1985}
is postponed to larger distances until the cell sizes are small enough
compared to the relevant scales. This significantly impacts the properties of the
hot wind (velocity, temperature, density etc., 
see \cref{fig:profiles_subsonic_5mdisk}) at all radii, but especially
in the inner CGM. It is difficult to provide a ``rule of thumb'' for
the resolution requirements. While the \citetalias{Chevalier1985}
solution provides a qualitative guide, its idealised features
(spherical symmetry, pure thermal energy injection, no central galaxy, 
no gravity, no cooling and no ambient medium) mean that it cannot
easily be used to make quantitative predictions in this context. The same is
true even for more sophisticated analytic models referenced in our
introduction.
However, based on our experiments, we find that the distance to the
outflow sonic point, $r_\mathrm{sonic}$, is a reasonable proxy for
the relevant scales.
We recommend that high specific energy components of winds should be resolved by
several resolution elements at a minimum i.e. $r_\mathrm{sonic}/r_\mathrm{cell}\gg1$. 
This constraint almost certainly has to be checked after the fact, since the location
of the sonic point and the density profile of the wind (on which the cell
size depends) are difficult to predict a priori for most scenarios
of sufficient complexity to be worthy of numerical study. A rough
estimate can be made along the lines that we described in \cref{sec:spatial_res}.

Despite the missing wind acceleration, we have demonstrated that
bulk mass and energy fluxes, as well as global SFRs, converge even
without the resolution boost in our simulations.
However, this may not always be the case
outside our particular idealised setup. If energy losses due to
radiative cooling or interaction with the CGM become significant,
the partitioning of energy becomes more important. 
Also,
despite our careful inclusion of a cooling flow component to
represent a realistic CGM, the complex nature of
accretion and mergers would likely lead to divergent behaviour
in a cosmological context between winds with such different
properties.
Furthermore, a failure to resolve the correct wind structure
in the inner CGM brings additional issues.
The evolution of cold material entrained in the hot wind will be sensitive
to its properties, particularly close to the galaxy (this interaction
requires even higher resolution to capture, but another part of
the \textsc{Arkenstone} model, to be published separately, treats this aspect).
Underestimating the hot wind velocity by a factor of $\sim2-10$ within
the inner 10~kpc (see \cref{fig:profiles_subsonic_5mdisk}) would make it
impossible to correctly follow the fate of embedded cold clumps, swept along
in the outflow.
Finally, of particular
relevance to the \textsc{Arkenstone} project is the effort to study the 
large-scale implications of results gained from small-scale simulations that
can resolve the generation of stellar feedback driven outflows from the ISM.
It is imperative that the evolution of the wind as it flows out into the CGM
is correctly resolved in order to validate such models by comparison to
observations.

While we have explored the difficulties of resolving high specific energy
flows in the context of the Lagrangian schemes (or pseudo-Lagrangian, in the
case of our adopted \textsc{Arepo} code), in practice these challenges
will also apply to adaptive mesh refinement (AMR) Eulerian codes deployed in
cosmological simulations. This is because these schemes typically also
reduce spatial resolution as density decreases (e.g. to resolve the Jeans
length by some number of cells or to target a particular cell mass range) to make
these simulations computationally feasible in the face of the large dynamic range
of scales associated with galaxy formation in a cosmological context.

Given the emphasis we have placed on the importance of using models
that can properly resolve these winds, an obvious question is whether
our \textsc{Arkenstone} scheme provides the correct result. Unfortunately,
as described above, there is no analytic solution to benchmark our simulations
against.
On the other hand, there is not much utility in applying \textsc{Arkenstone}
to an even more idealised setup for which an analytic solution exists.
The displacement recoupling scheme is designed specifically to solve
the issue of correctly injecting energy to the ISM/CGM transition region.
Wind particles have no use in a setup, such as that of \citetalias{Chevalier1985},
where there is neither an existing ISM or CGM. Indeed, we would
have to make such significant modifications
to our scheme to enable it to function in such a test
that any comparison to a benchmark solution is of little to no
value.
This motivated our choice
of the setup used in this paper; it is sufficiently simple to permit
insight from existing analytic solutions but contains enough of the features
of the target application of \textsc{Arkenstone} (cosmological simulations)
to represent a useful test. Nonetheless, we see that our results,
particularly the 
re-balancing of thermal to kinetic energy and the
radial profiles shown in \cref{fig:profiles_subsonic_5mdisk},
are in accordance with expectations from solutions such as \citetalias{Chevalier1985}.
Additionally, our refinement resolution study (\cref{sec:resolution}) demonstrates that these
results are robust with respect to resolution convergence.
Finally, it is worth re-emphasising that we have aimed to demonstrate in
this paper that our scheme is capable of resolving high specific energy
galactic winds where other approaches may fail, but \textit{not} to make any particular claim about
the most physical choice of mass and energy loadings. This will
be investigated in future work.

\subsection{The role of high and low specific energy winds in galaxy formation}
The mass and energy loadings used in this work were chosen to demonstrate some of the
features of the \textsc{Arkenstone} model, rather than being intended to precisely
represent the ``correct'' wind for our chosen galaxy. They also illustrate the
different ways in which winds can regulate galaxy properties. 
In recent years it has become increasingly evident that
a focus on the properties of the galaxies alone is
insufficient to discriminate between models. The properties of
halo gas may vary widely between different approaches that
all succeed in reproducing realistic galaxy stellar components
\citep[][]{Kelly2022}.
Our fiducial low and
high specific energy simulations have almost identical input energy loadings of $\sim0.9-1$,
high efficiencies being frequently adopted for systems of this mass 
($M_{200}=10^{11}\,\Msun$) in cosmological simulations.
The \lowspec{} simulation, with $\eta_M=6.41$ regulates
the galaxy by throwing large quantities of material out of the ISM. It does little
to arrest inflowing CGM material, but keeps the SFR low by keeping potential star
forming material circulating in a low altitude fountain flow. The \highspec{} and
\highark{} winds use approximately the same energy per unit SFR to throw a factor
twenty less mass. They are unable to significantly deplete the ISM, relative to
consumption by star formation, but the resulting hot, high velocity outflows
carve out a significant portion of the inflowing CGM, hindering inflows and
turning material around before it can reach the galaxy. In this way, as shown
in \cref{fig:cumul_star},
all simulations have formed the same stellar mass within the 2.5~Gyr of the
simulation (although the artificial start of the simulation and the lack
of cosmological context should be borne in mind).
However,
the resulting CGM properties are radically different. There is no longer a dense, cool
reservoir of mass being juggled close to the disc, replaced by a low density,
high temperature flow moving outwards at high velocity.
Both scenarios represent extreme cases, where almost all the available energy from
stellar feedback is coupled to a large or small quantity of gas.

Small scale
simulations that resolve the driving of winds out of the porous ISM
show that the energy is not monolithically coupled to a single component, but
is distributed unevenly to different elements of the multiphase outflow
\citep[e.g.][]{Kim2020b}. This allows both
low and high specific energy wind components to be driven simultaneously,
permitting both a mass loaded fountain flow and a hot, fast wind that
can impact the wider CGM. Cosmological simulations typically require
stellar feedback
driven winds with high mass loading factors to prevent 
runaway star formation 
and regulate other galaxy and CGM properties (particularly in sub-$L^\star$ galaxies).
This is potentially discrepant
with the lower mass loading factors that are increasingly being
found in simulations that can better resolve the generation of
winds \citep[see e.g.][]{Li2020a}. Likewise,
the high energy loading factors (sometimes super-unity)
frequently adopted are problematic when compared to 
simulations that resolve the multiphase star-forming ISM at $\sim$pc scales,
with the space-time correlations of star formation sites and feedback sites self-consistently determined.
It is possible that the division of mass and energy into high and low specific
energy components, providing both a long-range impact and a short-range
fountain flow, may alleviate some of these issues.
When we used a lower energy loading in \cref{sec:twindfix}
that is more compatible with predictions for the hot wind phase
\citep{Kim2020a}
the outflow was still able
to travel out beyond $r_{200}$ and impact the CGM structure
(as well as enriching it with metals, though this is beyond the
scope of this work). 
However,
it did not reduce the SFR compared to the no wind case as
much as our fiducial simulations.
Further suppression of SFR by winds may occur if we included
a more mass loaded, lower specific energy wind
component operating in tandem with the high specific energy
wind.
We will explore the interaction
of low and high specific energy winds in future \textsc{Arkenstone}
papers. We point out, however, that we used idealised non-cosmological
simulations in this work to allow an exploration of the behaviour of
our model, rather than to judge the ``success'' of any particular
set of loading factors. \textsc{Arkenstone} will be deployed in
cosmological simulations, where it will serve as a framework to
study the impact of small-scale models for galactic wind
generation and evolution on galaxy formation as a whole.

Finally, we remark that the current formulation of \textsc{Arkenstone}
makes no explicit inclusion of the impact of cosmic rays (CRs).
There is now significant theoretical evidence that CRs can
drive powerful galactic winds 
\citep[see e.g. section 3.4 of the review of][]{Naab2017} since
they can provide a non-thermal source of pressure with low
radiative losses. The effect of CRs on the launching
of winds out of the ISM can be trivially included in \textsc{Arkenstone}
in an implicit manner by
using input mass and energy loading factors derived from
small-scale simulations \citep[e.g.][]{Girichidis2016b,Simpson2016,Girichidis2018,Rathjen2021},
since by design our scheme is intended to model the evolution of
winds once they have left the ISM. Including CRs in the
evolution of the wind once wind particles have recoupled is
more complicated, since at that point the flow is treated
hydrodynamically rather than in a subgrid manner. 
Future work may include the extension of \textsc{Arkenstone}
to include an effective model, though this beyond the scope
of this paper.
\vspace{-4ex}
\section{Conclusions}
\textsc{Arkenstone} is a new model for the inclusion of multiphase
stellar feedback driven galactic winds in coarse resolution
cosmological hydrodynamic simulations of galaxy formation,
implemented in the \textsc{Arepo} code.
In this first presentation paper, we demonstrate aspects of
the model that allow it to treat high specific energy
(i.e. hot and fast) outflows. In particular, these include
novel schemes for achieving sufficient energy resolution
at the point of wind injection and for maintaining the
necessary spatial resolution required to capture the
subsequent wind evolution as it flows outwards.
For the purpose of clarity, we demonstrate
this subset of \textsc{Arkenstone}'s features with
non-cosmological simulations of a $M_{200}=10^{11}\,\Msun$
system with a cooling flow CGM, but at a resolution
that would be achievable in a simulation of a cosmological volume.
Our main findings are as follows:
\begin{enumerate}
\item Lagrangian hydrodynamic schemes tie energy resolution to mass
resolution at the point of wind injection. For high specific energy winds,
when combined with the coarse mass resolution needed for cosmological
volumes
this leads to an unphysically noisy coupling of feedback energy into
the gas.
We solve this issue by using higher resolution wind particles
and a new ``displacement recoupling'' scheme, which compensates for
the lack of a resolved, porous ISM/CGM transition region,
while avoiding numerical overcooling.

\item High specific energy winds have low densities. For a Lagrangian
code (or an Eulerian scheme with Lagrangian-like refinement), this
results in poor spatial resolution. We demonstrate that this can
lead to an incorrect partitioning of energy between kinetic and thermal
components in the wind. The wind velocity is underestimated and too much
thermal energy is retained.
Our novel refinement scheme ensures that
we resolve the sonic point of the flow, where the wind experiences
rapid acceleration.

\item As a qualitative demonstration, we compared a low specific energy wind,
with a high mass loading
factor comparable to that used in contemporary cosmological simulations,
to a high specific energy wind that had approximately the same energy loading
but a factor of twenty lower mass loading. We showed that both were able
to regulate the SFR of the galaxy, but by starkly contrasting mechanisms.
The mass loaded wind inhibited star formation by ejecting large quantities
of the ISM and keeping it circulating in a low altitude fountain flow. The
high specific energy wind did not eject the existing ISM, but inhibited
inflows through the entire halo such that the reservoir of gas available
for star formation was not re-supplied, leading to a gradual reduction in
the SFR.

\item When we used lower input mass and energy loading factors consistent with those measured
in high resolution simulations of the ISM for the hot component of multiphase outflows,
we found that the outflow travelled significantly into the CGM. However, inflows were
not disrupted sufficiently to reduce star formation to the levels
seen in our fiducial simulations. The inclusion of the lower specific energy
component (which is expected to dominate the mass loading
but is omitted in this demonstration) may lead to further reduction
in the SFR for this particular system.
\end{enumerate}

Subsequent work will present the remaining aspects of the \textsc{Arkenstone} model
not used in this paper.
One important feature is the simultaneous injection of low and high specific energy wind 
components with velocities and temperatures drawn from distributions. These distributions
are derived from measurements made from small scale simulations that resolve the generation
of the outflows from within the ISM. The other key component of the \textsc{Arkenstone} model
is the use of ``cloud particles'' to treat the unresolvable interactions between
the hot wind phase and an embedded population of entrained cold clouds. This is necessary
to properly capture the evolution of a multiphase wind. The \textsc{Arkenstone} model
will be used as a framework to study the implications of models of small scale physics 
(e.g. the formation of galactic winds, cold cloud acceleration and shredding) in the wider context of
galaxy formation. Finally, the model will be deployed in large volume cosmological
simulations.

\section*{Acknowledgements}
We are grateful to Max Gronke and Annalisa Pillepich for helpful
comments and discussions.
This work was carried out as part of the SMAUG project. SMAUG gratefully acknowledges support from the Center for Computational Astrophysics at the Flatiron Institute, which is supported by the Simons Foundation.
This work was supported by the Simons Collaboration on “Learning the Universe.”
The work of MCS was supported by a grant from the Simons Foundation (CCA 668771, LEH) and by the DFG under Germany’s Excellence Strategy EXC 2181/1-390900948 (the Heidelberg STRUCTURES Excellence Cluster).
GLB acknowledges support from the NSF (AST-2108470, XSEDE grant MCA06N030), NASA TCAN award 80NSSC21K1053, and the Simons Foundation (grant 822237).
ECO and C-GK were supported by the Simons Foundation (grant 888968).
JS was supported by the Israel Science Foundation (grant No. 2584/21).
KS acknowledges support from the Black Hole Initiative at Harvard University, which is funded by grants from the John Templeton Foundation and the Gordon and Betty Moore Foundation.
RW was supported by the Natural Sciences and Engineering Research Council of Canada (NSERC), funding reference \#CITA 490888-16.
CYH acknowledges support from the DFG via German-Israel Project Cooperation grant STE1869/2-1 GE625/17-1.
BB is grateful for the generous support from the David and Lucile Packard Foundation and Alfred P. Sloan Foundation.
YL acknowledges financial support from NSF grants AST-2107735 and AST-2219686, and NASA grant 80NSSC22K0668. 
Computations were performed on the HPC system Raven at the Max Planck Computing and Data Facility (MPCDF).
The following open source software packages were used in this work:
\texttt{Astropy} \citep{AstropyCollaboration2013,AstropyCollaboration2018,AstropyCollaboration2022},
\texttt{Matplotlib} \citep{Hunter2007},
\texttt{nanoflann} \citep{Blanco2014},
\texttt{NumPy} \citep{Harris2020},
\texttt{SciPy} \citep{Virtanen2020}.

%%%%%%%%%%%%%%%%%%%%%%%%%%%%%%%%%%%%%%%%%%%%%%%%%%
\section*{Data Availability}

The data underlying this article will be shared on reasonable request to the corresponding author.



\bibliographystyle{mnras}
\bibliography{references}

%%%%%%%%%%%%%%%%%%%%%%%%%%%%%%%%%%%%%%%%%%%%%%%%%%

%%%%%%%%%%%%%%%%% APPENDICES %%%%%%%%%%%%%%%%%%%%%

\appendix

\section{Host cell mass retention fraction} \label{ap:fret}
As described in \cref{sec:disp_rec}, under some circumstances it is possible that
after a wind particle displacement recouples into a host cell, the resulting
hot wind cell will be under-pressured relative to the ambient medium. This can
occur if the temperature contrast between the wind particle and the ambient medium is small
\textit{and} the wind particle has a small mass compared to the mass
resolution
of cells in the ambient medium.

Under the simplifying assumption that the ambient medium is homogeneous, the pressure
contrast between the host cell and a neighbour cell after recoupling is:
\begin{equation} \label{eq:pressure_contrast_1}
    \chi_P = \frac{P_\mathrm{host}}{P_\mathrm{ngb}} = \frac{\rho_\mathrm{host}u_\mathrm{hot}}{\rho_\mathrm{ngb}u_\mathrm{ngb}} = \frac{E_\mathrm{th,host}}{E_\mathrm{th,ngb}},
\end{equation}
where we have additionally assumed all cells have the same mass prior to recoupling (and hence the same volume before and after recoupling)
to obtain the last equivalence.

We specify a minimum pressure contrast, $\chi_{P,\mathrm{min}}$, that we wish to achieve after the displacement recoupling is complete. Our fiducial
choice is $\chi_{P,\mathrm{min}}=1.1$ which should prevent an initial inflow of cold material back into the host cell. However, as explained above,
this pressure contrast may not be met if all of the material is displaced. We can therefore compromise and retain a fraction of the original host
cell material, $f_\mathrm{ret}$, such that the desired pressure contrast is achieved at the cost of a cooler cell. Maintaining the
approximation that all cells were homogeneous before recoupling, recognising that displaced material conserves specific thermal energy,\footnote{We do not increase the internal energy of neighbour cells to account for adiabatic
compression. This increase would be very small (since neighbour cells typically receive a small fraction each of the
original host cell material) and we assume that it is radiated away. Our tests show that including this heating source has no detectable impact on our results. Likewise, we do not decrease the thermal energy of wind particles to account for work done compressing displaced material as we assume these losses are already included in the input energy loadings.}
considering the pressure contrast with respect to the cells receiving displaced material and assuming that all receive the same fraction of material
(i.e. ignoring the kernel weighting)
we can derive an approximate expression for the smallest retention fraction that will achieve the desired minimum pressure contrast:
\begin{equation} \label{eq:retention_fraction}
    f_\mathrm{ret} = \mathrm{MAX}\left[\frac{E_\mathrm{th,host,0}\chi_{P,\mathrm{min}}\left(1 + \frac{1}{N_\mathrm{ngb,el}}\right) - u_\mathrm{w}m_\mathrm{w}}{E_\mathrm{th,host,0}\left(1 + \frac{\chi_{P,\mathrm{min}}}{N_\mathrm{ngb,el}}\right)},0\right],
\end{equation}
where $E_\mathrm{th,host,0}$ is the thermal energy of the host cell (and its neighbours) prior to recoupling and
$N_\mathrm{ngb,el}$ is the number of eligible neighbours.
The simplifying assumptions of homogeneity could be relaxed by explicitly checking the values of neighbouring cells, but this would require
a more complicated implementation. \Cref{eq:retention_fraction} has the advantage of being a reasonable guess that can be evaluated
locally. Note that this approximation neglects relative velocities between cells prior to recoupling. Likewise, it does not consider whether
any portion of the wind particle's kinetic energy will be thermalised upon recoupling. It therefore should be treated as a conservative estimate of
the required retention fraction, which suits our purposes.

If $f_\mathrm{ret} > 1$, the desired minimum overpressure is unachievable, in which case we fall back to a standard recoupling. In practice,
for the wind loadings and CGM initial conditions presented in this work, only a negligible fraction of recouplings require any mass to be
retained in order to achieve $\chi_P = 1.1$.

\section{Convergence with base resolution} \label{ap:base_res}
\begin{figure} 
\centering
\includegraphics{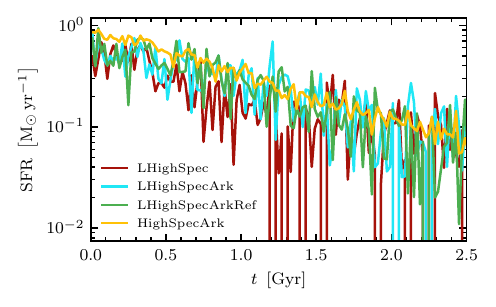}
\caption{Star formation rate as a function of time
for the fiducial ICs with the base mass resolution of the simulation (gas cells,
star and wind particles) coarsened by a factor of 8. 
Other than the coarser resolution, \lhighspec{} and \lhighark{} are
identical in setup to their fiducial resolution equivalents. \lhigharkref{}
uses a factor of 8 smaller value of $f_{m,\mathrm{w}}$, compensating such
that the resolution in the hot wind is the same as the fiducial \highark{}
simulation. The latter is also shown for reference.
The coarser resolution simulations have marginally
lower SFRs at a given time and are more bursty.}
\label{fig:sfr_subsonic_5mdisk_base_resolution} 
\end{figure}

\begin{figure*} 
\centering
\includegraphics{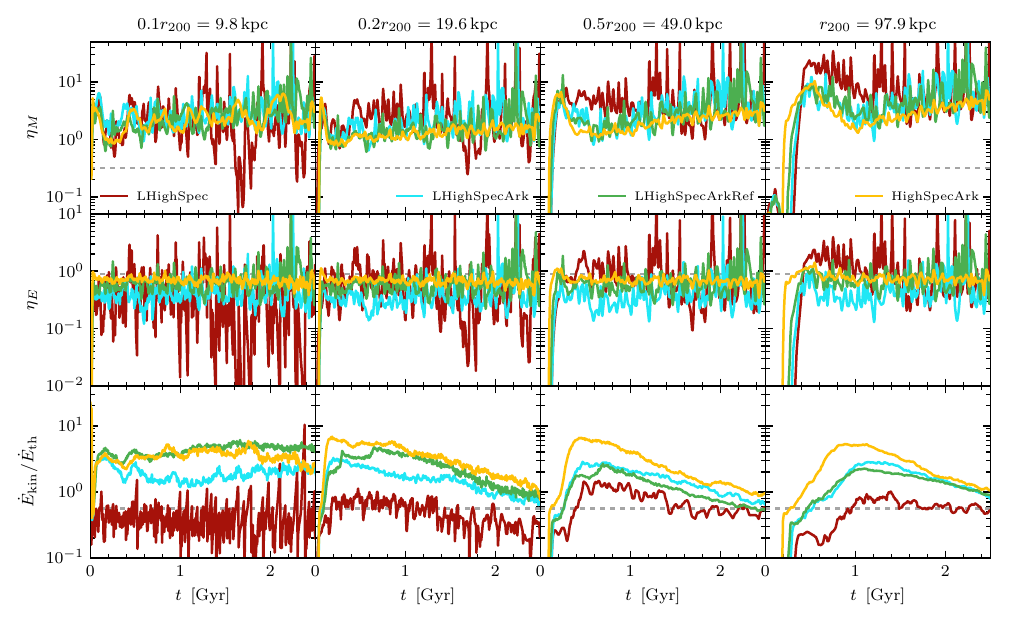}
\caption{Mass loadings (top row), energy loadings (middle row)
and ratio of kinetic to thermal energy fluxes (bottom row)
for the fiducial ICs with the base mass resolution of the simulation (gas cells,
star and wind particles) coarsened by a factor of 8. 
Other than the coarser resolution, \lhighspec{} and \lhighark{} are
identical in setup to their fiducial resolution equivalents. \lhigharkref{}
uses a factor of 8 smaller value of $f_{m,\mathrm{w}}$, compensating such
that the resolution in the hot wind is the same as the fiducial \highark{}
simulation. The latter is also shown for reference. The
horizontal dashed lines show the input loadings and the resulting input kinetic to thermal ratio.
The \textsc{Arkenstone} model continues to work at this lower resolution, smoothing out the
artificial burstiness of the outflow rates and capturing the conversion of thermal to kinetic
energy as the wind flows outwards.
}
\label{fig:load_relative_subsonic_5mdisk_base_resolution} 
\end{figure*}

In \cref{sec:resolution} we demonstrated the impact of varying the wind refinement factor, $f_{m,\mathrm{w}}$,
on the evolution of the wind. Here, we perform a similar exercise, but coarsen the base resolution of the
entire simulation. We use the same structural parameters as the fiducial ICs described in
\cref{tab:IC}, but increase $m_\star$ and $m_\mathrm{g,tar}$ by a factor of 8 to $6.4\times10^5\,\mathrm{M_\odot}$.
Gravitational softening lengths are likewise increased by a factor of 2. We carry out three simulations with
these coarse resolution ICs. \lhighspec{} and \lhighark{} are the coarser resolution equivalents to
\highspec{} and \highark{}, respectively.
\lhigharkref{} is a variant of \lhighark{} with
$f_{m,\mathrm{w}}$ decreased from our fiducial value of 0.01 to 0.00125. In the latter case, 
the decreased $f_{m,\mathrm{w}}$ compensates such that the wind particle mass and the mass resolution at the base
of the wind match those in the fiducial \highark{} simulation, despite the coarser target gas mass elsewhere.

In \cref{fig:sfr_subsonic_5mdisk_base_resolution} we show SFRs as a function of time, calculated from the mass
in new star particles created over the preceding 20~Myr, for these simulations.
For reference, we also show the fiducial resolution \highark{} run.
All simulations show similar
behaviours, with the SFR declining from its initial value as the gas supply is used up faster than it is replenished,
due to the preventative feedback of the winds.
The fiducial \highark{} run has a marginally enhanced SFR at any given time compared to the coarse resolution
reruns. This is because with higher resolution, gas in the ISM is able to reach higher densities
\citep[see e.g. the discussion in][appendix A]{Pillepich2018}.
The rates are also burstier in the coarse resolution runs. This is mainly due to the increased Poisson noise associated
with stochastically forming more massive star particles, although the increased burstiness of the winds (as we shall show
next) plays a role.

In \cref{fig:load_relative_subsonic_5mdisk_base_resolution} we show the emergent mass and energy loading factors
of the winds (measured in the same way as in the rest of this work), as well as the ratio of kinetic to thermal
energy fluxes. The colours are the same as \cref{fig:sfr_subsonic_5mdisk_base_resolution}. The results are
broadly similar to those presented in \cref{sec:resolution}. The mass and energy loadings are similar
between the four simulations. However, the \lhighspec{} simulation (which does not benefit from the new
\textsc{Arkenstone} techniques) is even more bursty than its fiducial resolution counterpart
(see \cref{fig:load_relative_subsonic_5mdisk}) because it suffers even more severely from the Poisson noise
described in \cref{sec:challenge}. The \lhighark{} simulation significantly suppresses this artificial burstiness.
In \lhigharkref{}, reducing $f_{m,\mathrm{w}}$ results in marginally smoother outflow rates, as expected. The fiducial resolution
\highark{} simulation has the smoothest outflow rates because, in addition to the higher resolution wind particles (the same
as \lhigharkref{}), it has more resolution elements in the ISM and the material inflowing to the disc, so the underlying SFR
driving the wind is less noisy.

When examining the ratio of kinetic to thermal energy in the wind fluxes, we see a large contrast between the simulations
that use the full \textsc{Arkenstone} scheme and \lhighspec{}. The latter simulation does not capture the conversion
of thermal to kinetic energy as the wind flows outwards, due to failing to resolve the sonic point of the wind
to an even greater degree than the fiducial \highspec{} run. When comparing \lhighark{} and \highark{} (which 
differ only in base resolution of the simulation but otherwise have the same model parameters), we see that
the former has a marginally lower ratio of energy components. This suggests that the wind is not quite as well
resolved. The offset is similar to that seen in \cref{fig:load_relative_subsonic_5mdisk_resolution} where we
kept the base resolution the same but increased $f_{m,\mathrm{w}}$ by a factor of 10. When we increase the
resolution in the wind, \lhigharkref{}, so that it matches \highark{}, this small difference is largely
removed for the measurements at 0.1$r_{200}$ and 0.2$r_{200}$. These results demonstrate that it is
largely the absolute resolution in the wind that is important, as discussed in \cref{sec:resolution}, rather
than the base resolution of the entire simulation. By 0.5$r_{200}$ and beyond, the ratio has dropped
slightly to be closer to the values seen in \lhighark{}; at this radius the two simulations have the
same resolution. The wind is therefore not completely converged with respect to the base resolution
in the outer halo, where the resolution of the ambient and inflowing material likely plays a
role in the interaction with and confinement of the outflow.

We have demonstrated that the scheme functions well for a lower resolution ($m_\mathrm{g,tar}=6.4\times10^5\,\mathrm{M_\odot}$)
than that used in the main body of this work. If the resolution were coarsened by a further factor of 8, while keeping the
value of $f_{m,\mathrm{w}}$ at our fiducial choice of 0.01, then it would likely fail for \emph{this} set of
wind parameters. At that point the mass resolution of the wind particles and hot wind cells would approach
that used in our fiducial \highspec{} simulation (which did not use refinement). However, we have demonstrated
here and in \cref{sec:resolution} that additional refinement can be used to compensate for this. We also stress
again that the required resolution is heavily dependent on the nature of the wind. In this case, we are using
a very high specific energy wind as a test case. We therefore leave further demonstrations of the feasibility of
the scheme at coarser resolutions until a future work in a cosmological context, where we will also be able
to more properly characterise the computational cost of reducing $f_{m,\mathrm{w}}$ further.

%%%%%%%%%%%%%%%%%%%%%%%%%%%%%%%%%%%%%%%%%%%%%%%%%%

% Don't change these lines
\bsp	% typesetting comment
\label{lastpage}
\end{document}